\documentclass[aps,pra,twocolumn,groupedaddress,longbibliography,notitlepage=,
floatfix,superscriptaddress]{revtex4-2}

\pdfoutput=1
\usepackage{graphicx,graphics,epsfig,subfigure,times,bm,bbm,amssymb,amsmath,amsthm,mathrsfs,MnSymbol}
\usepackage{gensymb}
\usepackage{amsfonts}
\usepackage{float}
\usepackage{ragged2e}
\usepackage[matrix,frame,arrow]{xypic}
\usepackage[pdfstartview=FitH]{hyperref}
\hypersetup{
	colorlinks=true,       
	linkcolor=red,          
	citecolor=blue,        
	filecolor=magenta,      
	urlcolor=blue,           
	runcolor=cyan
}
\usepackage[pdftex]{color}
\usepackage{braket}  
\usepackage{enumerate}
\usepackage[normalem]{ulem}
\usepackage[usenames,dvipsnames]{xcolor}
\usepackage{multirow}
\usepackage{mathtools}
\usepackage{bbm}
\usepackage{titletoc}

\definecolor{orange}{rgb}{1,0.5,0}

\newcommand{\RNum}[1]{\uppercase\expandafter{\romannumeral #1\relax}}

\newcommand{\ignore}[1]{}
\usepackage{geometry}

\newcommand{\x}{\mathbf{x}}

\begin{document}
\title{Thermal variational quantum simulation on a superconducting quantum processor}

	\author{Xue-Yi~Guo}
	\altaffiliation[]{These authors contributed equally to this work.}
	\affiliation{Beijing National Laboratory for Condensed Matter Physics, Institute of Physics, Chinese Academy of Sciences, Beijing 100190, China}
	\affiliation{School of Physical Sciences, University of Chinese Academy of Sciences, Beijing 100190, China}
	
	\author{Shang-Shu~Li}
	\altaffiliation[]{These authors contributed equally to this work.}
	\affiliation{Beijing National Laboratory for Condensed Matter Physics, Institute of Physics, Chinese Academy of Sciences, Beijing 100190, China}
	\affiliation{School of Physical Sciences, University of Chinese Academy of Sciences, Beijing 100190, China}
	
	\author{Xiao~Xiao}
	\altaffiliation[]{These authors contributed equally to this work.}
		\affiliation{The Chinese University of Hong Kong, Shatin, New Territories, Hong Kong, China}

	  \author{Zhong-Cheng Xiang}
	\affiliation{Beijing National Laboratory for Condensed Matter Physics, Institute of Physics, Chinese Academy of Sciences, Beijing 100190, China}
	 \affiliation{School of Physical Sciences, University of Chinese Academy of Sciences, Beijing 100190, China}
	 
    \author{Zi-Yong Ge}
	\affiliation{Beijing National Laboratory for Condensed Matter Physics, Institute of Physics, Chinese Academy of Sciences, Beijing 100190, China}
    \affiliation{School of Physical Sciences, University of Chinese Academy of Sciences, Beijing 100190, China}
    
    \author{He-Kang Li}
	\affiliation{Beijing National Laboratory for Condensed Matter Physics, Institute of Physics, Chinese Academy of Sciences, Beijing 100190, China}
    \affiliation{School of Physical Sciences, University of Chinese Academy of Sciences, Beijing 100190, China}

    \author{Peng-Tao Song}
	\affiliation{Beijing National Laboratory for Condensed Matter Physics, Institute of Physics, Chinese Academy of Sciences, Beijing 100190, China}
	 \affiliation{School of Physical Sciences, University of Chinese Academy of Sciences, Beijing 100190, China}
	\author{Yi Peng}
	\affiliation{Beijing National Laboratory for Condensed Matter Physics, Institute of Physics, Chinese Academy of Sciences, Beijing 100190, China}
	 \affiliation{School of Physical Sciences, University of Chinese Academy of Sciences, Beijing 100190, China}
	
	\author{Kai~Xu}
	\affiliation{Beijing National Laboratory for Condensed Matter Physics, Institute of Physics, Chinese Academy of Sciences, Beijing 100190, China}
	\affiliation{School of Physical Sciences, University of Chinese Academy of Sciences, Beijing 100190, China}
	\affiliation{CAS Center for Excellence in Topological Quantum Computation, UCAS, Beijing 100190, China}

	\author{Pan~Zhang}
	\email{panzhang@itp.ac.cn}
	\affiliation{
    Institute of Theoretical Physics, Chinese Academy of Sciences, Beijing 100190, China
}
\affiliation{School of Fundamental Physics and Mathematical Sciences, Hangzhou Institute for Advanced Study, UCAS, Hangzhou 310024, China}
\affiliation{International Centre for Theoretical Physics Asia-Pacific, Beijing/Hangzhou, China}
	
	\author{Lei~Wang}
	    \email{wanglei@iphy.ac.cn}
		\affiliation{Beijing National Laboratory for Condensed Matter Physics, Institute of Physics, Chinese Academy of Sciences, Beijing 100190, China}
    \affiliation{Songshan Lake Materials Laboratory, Dongguan, Guangdong 523808, China}
	
		\author{Dong-Ning Zheng}
	   \email{dzheng@iphy.ac.cn}
		\affiliation{Beijing National Laboratory for Condensed Matter Physics, Institute of Physics, Chinese Academy of Sciences, Beijing 100190, China}
	\affiliation{School of Physical Sciences, University of Chinese Academy of Sciences, Beijing 100190, China}
	
	\author{Heng~Fan}
	\email{hfan@iphy.ac.cn}
	\affiliation{Beijing National Laboratory for Condensed Matter Physics, Institute of Physics, Chinese Academy of Sciences, Beijing 100190, China}
	\affiliation{School of Physical Sciences, University of Chinese Academy of Sciences, Beijing 100190, China}
	\affiliation{CAS Center for Excellence in Topological Quantum Computation, UCAS, Beijing 100190, China}
	\affiliation{Beijing Academy of Quantum Information Sciences, Beijing 100193, China}

	\begin{abstract}
    Solving finite-temperature properties of quantum many-body systems is generally challenging to classical computers due to their high computational complexities.
   In this article, we present experiments to demonstrate a hybrid quantum-classical simulation of thermal quantum states.
   By combining a classical probabilistic model and a 5-qubit programmable superconducting quantum processor, we prepare Gibbs states and excited states of Heisenberg XY and XXZ models with high fidelity and compute thermal properties including the variational free energy, energy, and entropy with a small statistical error. Our approach combines the advantage of classical probabilistic models for sampling and quantum co-processors for unitary transformations. We show that the approach is scalable in the number of qubits, and has a self-verifiable feature, revealing its potentials in solving large-scale quantum statistical mechanics problems on near-term intermediate-scale quantum computers. 

\end{abstract}
\maketitle{\tiny}

\begin{figure*}[!tp]
	\includegraphics[width=0.8\textwidth]{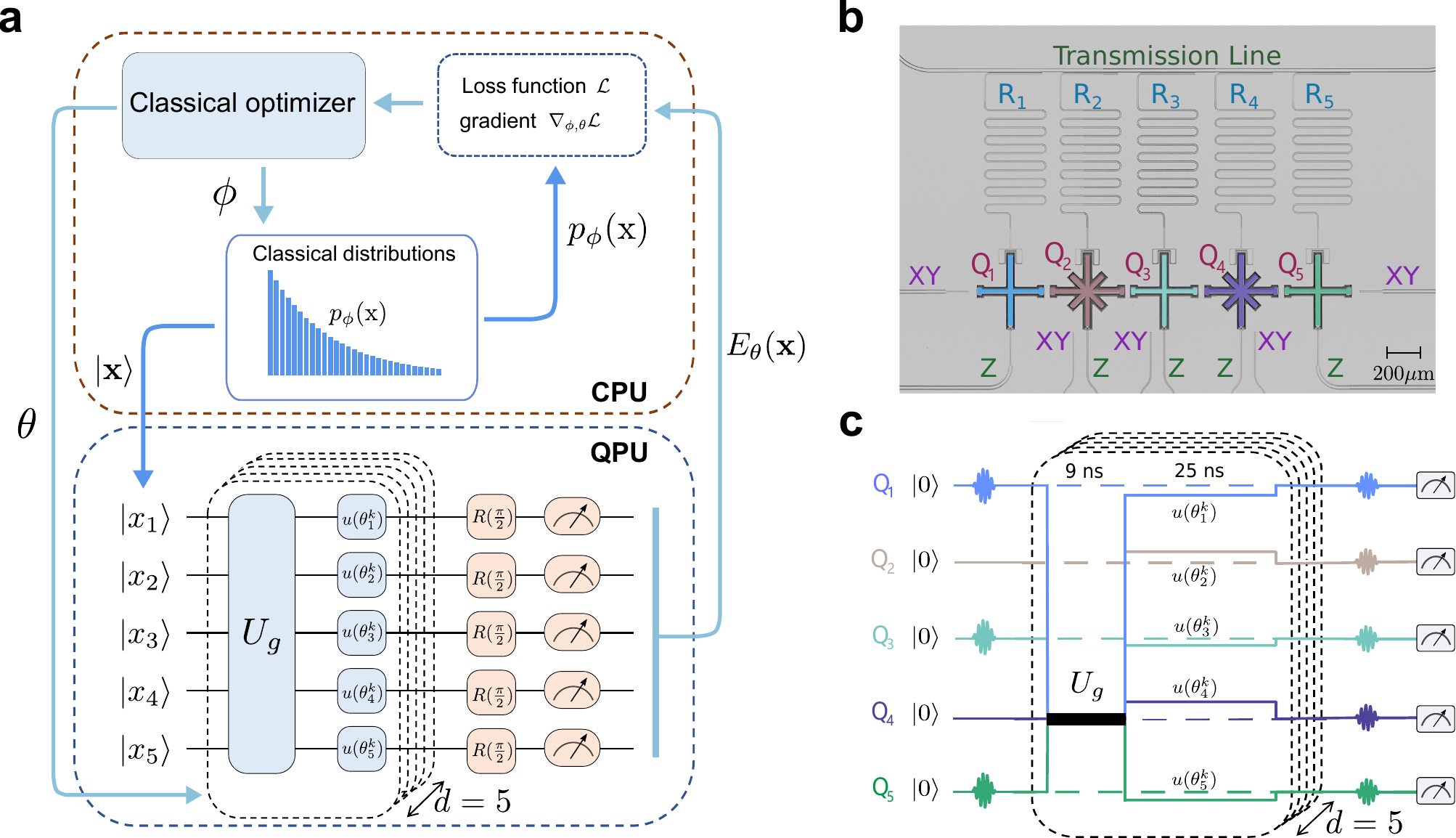}
	\caption{
	\textbf{Thermal vaiational quantum simulaiton on a superconducting quantum processor.} 
	\textbf{a,} Feedback loop of the hybrid quantum-classical variational algorithm in our experiment between the classical process unit (CPU) and the quantum process unit (QPU). 
	A classical probability distribution $p_\phi(\x)$ is maintained using classical computational resources, produces bitstring samples $\{\x\}$ which work as the input product states $|\mathbf{x}\rangle$ for QPU. 
	The QPU performs several layers of unitary quantum circuits $U(\mathbf{\theta})$ with the input product states and produces the final states by which the energy $E_{\theta}(\mathbf{x})$ is estimated. The energy is forwarded to the CPU for evaluating the loss function and its gradients for the parameters of $p_\phi$ and $U_\theta$. Then a classical optimizer is employed to update the parameters upon which new bitstring samples are generated for the next loop until the loss function converges.
	\textbf{b,} False-color image of our quantum device. There are 5 frequency-tunable transmon qubits lay at the middle of the chip, each of them has a readout resonator, an XY control line, and a Z control line. All the 5 readout resonators are coupled to a readout transmission line. See Appendix~\ref{app_D} for the  basic  characteristics of qubits (Table \ref{tab1}). 
	{\textbf{c,}} Pulse sequence for the quantum circuit ansatz for a certain input state ($|x\rangle = |10101\rangle$). The evolution of the global entanglement layer (with the XY couplings) persists 9ns. The amplitude tunable square pulses of 25ns are used to realize the parameterized $R_z(\theta)$. After $d$-layer evolutions ($d=5$ in our experiment), the $-X/2$, $Y/2$, or $I$ gates are applied for each qubit for then energy measurement.
	}
	\label{fig:fig1}
\end{figure*}

Investigating quantum statistical mechanics is challenging for classical computations, due to the exponential growth of the Hilbert space dimension. Conventional approaches suffer from the fundamental difficulties in sampling, approximating the density matrix, and computing the partition function. The recently developed quantum devices are promising to resolve the difficulties by utilizing quantum resources. 
Particularly, the hybrid quantum-classical algorithms take advantage of both classical algorithms and currently available quantum resources, by arranging the classically high-cost computational part to the quantum co-processors. Such approaches are feasible on near-term noisy intermediate-scale quantum (NISQ) devices for robustness in the presence of noise and decoherence~\cite{preskill_quantum_2018, moll_quantum_2018}, and have been demonstrated for solving combinatorial optimization problems with quantum approximate optimization algorithm~\cite{farhi_quantum_2014, zhou_quantum_2020}, the variational quantum time evolution of quantum systems~\cite{mcardle_variational_2019, endo_variational_2020}, and studying the ground states in various systems with the variational quantum eigenstate solver (VQE)~\cite{peruzzo_variational_2014, colless_computation_2018,collaborators_hartree-fock_2020,hempel_quantum_2018,kandala_hardware-efficient_2017,omalley_scalable_2016,kandala_hardware-efficient_2017,kokail_self-verifying_2019}.

In quantum statistical mechanics, both the ground states and excited states contribute to the Gibbs states in thermal equilibrium. 
 Studying the Gibbs state is generally more difficult than the ground state for a quantum computer that can only provide pure states, as it is not straightforward to prepare the excited states and mix them according to the Boltzmann distribution.
Recently, some efforts have been made in preparing Gibbs states~\cite{terhal_problem_2000, poulin_sampling_2009, temme_quantum_2011,brandao_finite_2019, wu_variational_2019, zhu_generation_2020, sagastizabal_variational_2020,wang_variational_2020, motta_determining_2020, sun_quantum_2021, verdon_quantum_2019, liu_solving_2021} and excited states~\cite{santagati_witnessing_2018, higgott_variational_2019, jones_variational_2019, nakanishi_subspace-search_2019} using quantum computers, with experimental demonstrations~\cite{sagastizabal_variational_2020, sun_quantum_2021}. However, to perform these protocols on the NISQ devices, the existing approaches have applicability and scalability issues. 
For example, approaches accessing thermal Gibbs ensemble by variationally preparing thermal field double states~\cite{wu_variational_2019,sagastizabal_variational_2020} require an approximate measurement of entropy, which takes more experimental resources.
Quantum imaginary time evolution suffers from an inefficiency with deep circuits at low temperature or large system size~\cite{motta_determining_2020, sun_quantum_2021}. Moreover, it is also challenging to extract physical observables from an exponential large mixed state in a scalable way~\cite{sun_quantum_2021}.

In this article, we present a hardware implementation of a general variational algorithm for quantum statistical mechanics problems~\cite{liu_solving_2021, martyn_product_2019, verdon_quantum_2019}. In the approach, the variational thermal state is constructed by generating a mixture of product states using a classical probability distribution, as input states to a quantum circuit that perform unitary transformations and induce entanglements in the output states. It can be regarded as a finite temperature generalization of VQE, with initial states sampled from a classical distribution. Compared with other algorithms for thermal state preparations~\cite{sagastizabal_variational_2020,wang_variational_2020, motta_determining_2020, sun_quantum_2021}, the classical representation of the mixture probabilities reduces the burden of the quantum processor, has an advantage of efficiently estimating the entropy with small or even no statistical error, and benefits from the methods and models rapidly developed in machine learning. Consequently, the combined approach is scalable and particularly feasible for NISQ implementations.

Here we report our implementation of this approach on a 5-qubit superconductor quantum processor. We prepare Gibbs states and thermal states for quantum XY chain and XXZ chain using a symmetry-preserved analog-digital hybrid quantum ansatz.
The classical distribution and quantum circuit are trained to minimize the variational free energy via a mini-batch gradient descent algorithm which requires a small number of samples in each training step. The error of the estimated free energy is 5\% to the exact value. We show that the approach has a self-verification property~\cite{kokail_self-verifying_2019} and can be scaled up to a large system. We further evaluate the performance of our approach using state tomography, where the target Gibbs state can be constructed with fidelity reaching 92.6\%. Moreover, we illustrate that the classical probability can help us prepare specified eigenstates, of which the highest-fidelity reaches 98\%. Finally, we give a scalable approach based on thermodynamics relation to reduce statistical error in the calculation of some thermal variables.

\section{The hybrid quantum-classical variational approach}

In general, the thermal density matrix of a target quantum system can be represented as a classical mixture of pure states. The pure states can be prepared by applying a parameterized unitary quantum circuit $\hat U_\theta$ on a set of input state \{$\ket{x}$\}, and the classical mixture can be realized by using probabilistic generative model. This gives an hybrid ansatz~\cite{verdon_quantum_2019, liu_solving_2021} for the thermal density matrix,
\begin{equation}\label{ansatz}
	\hat{\rho}=\sum_x p_\phi(\x) \hat{U}_{\theta}\ket{\x}\bra{\x}\hat{U}_{\theta}^{\dag},
\end{equation}
where the set \{$\ket{x}$\} denotes the computational basis, $p_\phi(\x)$ is the parameterized probability distribution model satisfying $\sum_x p_\phi(\x)=1$ and $U_\theta$ is the variational quantum circuit.
The variational parameters $\theta, \phi$ are determined by minimizing the distance between $\hat{\rho}$ and target density matrix $\hat{\sigma}=e^{-\beta \hat{H}}/\mathcal{Z}$,
 where $\hat H$ is the Hamiltonian of the target quantum system, $\beta$ denotes the inverse temperature $\beta$, and $\mathcal{Z}=\text{Tr}(e^{-\beta \hat{H}})$ is the partition function. 
The Gibbs-Delbr\"uck-Moli\'eve variational principle of quantum statistical mechanics~\cite{huber1968variational} suggests to take the variational free energy
$\mathcal{L}=\frac{1}{\beta}\text{Tr}(\hat{\rho} \ln \hat{\rho}) + \text{Tr}( \hat{\rho} \hat{H})$ as the loss function, which is lower bounded by the true free energy with equality holds only when $\hat{\rho}=\hat{\sigma}$. 

Using the variational ansatz in Eq.~(\ref{ansatz}), the loss function is written as
\begin{equation}\label{loss}
	\mathcal{L}=\mathbb{E}_{x \sim p_{\phi} (\x)} [\frac{1}{\beta}\ln p_{\phi} (\x) + 
	\langle \psi_{\mathbf{\theta}}(\x)|\hat{H} |\psi_{\mathbf{\theta}}(\x)\rangle],
\end{equation}
where we have defined $|\psi_{\mathbf{\theta}}(\x)\rangle=\hat{U}(\mathbf{\theta})|\x\rangle$.
The loss function can be separated into the energy term and the entropy term.
The energy evaluation $\mathbb{E}_{x \sim p_{\phi} (\x)} [\langle \psi_{\mathbf{\theta}}(\x)|\hat{H} |\psi_{\mathbf{\theta}}(\x)\rangle]$ can be performed efficiently with the help of a quantum processor for preparing $|\psi_\theta(\x)\rangle$. While for the entropy term $\mathbb{E}_{x \sim p_{\phi} (\x)} [\ln p_{\phi}(\x)]$ which is difficult to estimate using quantum computer, it now can be evaluated pure classically based on a proper probabilistic model on binary variable $\x$. 
There are two ways to obtain the expectation in Eq.~(\ref{loss}) which corresponds to the thermal average. When the number of qubits $N$ is small, $p_\phi(x)$ can be exactly characterized by storing probabilities for totally $2^N$ computational basis vectors, and the thermal average is done using all the basis and corresponding probabilities. We term this as the \textit{full-space method}. Apparently, the computational cost of this approach is exponential in $N$. Another scalable way is to represent the probability distribution by a parameterized generative model and evaluate the thermal observables with sample mean, which we term as \textit{sample method}.

Here we use the \textit{sample method} combined with a classical gradient-based optimizer to optimize the hybrid variational model in a batch gradient descent manner (see Appendix~\ref{app_A}).
At each step of the optimization, a set of bitstrings $\{\x\}$ are sampled from $p_{\phi}(\x)$ generating initial states $|\x\rangle$ as inputs to the quantum circuit $U_\theta$.  After applying unitary transformations, the outputs of the quantum circuit work as final states $|\psi_{\mathbf{\theta}}(\x)\rangle$, using which we measure energy $E_{\mathbf{\theta}}(\x)=\langle\psi_{\mathbf{\theta}}(\x)|\hat{H}|\psi_{\mathbf{\theta}}(\x)\rangle$. 
Together with the entropy estimated solely using $p_\phi(\x)$, we compute the loss function and its gradients (see Appendix~\ref{app_A}), then employ an efficient classical optimizer (Appendix~\ref{app_A}) to update parameters.
The overall classical-quantum optimization process is sketched in Fig.~\ref{fig:fig1}a.

\section{Experimental realizations}

We focus on the spin-1/2 XXZ Heisenberg chain in a magnetic field, which is a prototype model studied extensively in condensed matter physics and has a rich phase diagram at finite temperature. The Hamiltonian is written as
\begin{equation}\label{Hamil}
\hat{H}_{T} = \sum_i \hat{\sigma}^x_i \hat{\sigma}^x_{i+1} + \hat{\sigma}^y_i \hat{\sigma}^y_{i+1}+\Delta \hat{\sigma}^z_i \hat{\sigma}^z_{i+1} + h\sum_i \hat{\sigma}^z_i.
\end{equation}
where $\hat{\sigma}^{x,y,z}$ are Pauli matrices. The model has a global $U(1)$ symmetry corresponding to the conservation of total spin on the $z$ direction. 
In the limit with $\Delta=0$, the model reduces to the XY model, which can be mapped to the free-fermion model by Jordan-Wigner transformation thus exactly solvable.  

Our experiments build upon a quantum processor with 5 frequency-tunable transmon qubits \cite{barends_coherent_2013} arranged in a line with nearest-neighbor couplings, whose false-color image is presented in Fig.\ref{fig:fig1}b. The qubit chain can be modeled as a one-dimensional spin-1/2 model with XY interactions, of which the Hamiltonian is expressed as
\begin{equation}\label{expH}
	\hat{H} = \sum_i^{N-1} g_{i, i+1}(\hat{\sigma}^{+}_i \hat{\sigma}^{-}_{i+1} + \hat{\sigma}^{-}_i \hat{\sigma}^{+}_{i+1}) + \sum_i^{N} \frac{\Delta_i}{2} \hat{\sigma}^z_i,
\end{equation}	
where $\hat{\sigma}^{\pm}=(\hat{\sigma}^x\pm i\hat{\sigma}^y)/2$ represent the spin raising and lowering operators respectively, $g_{i,i+1}$ is the spin coupling strength.
The architecture can easily realize the unitary operations  generated directly by the Hamiltonian~\eqref{expH}, as well as single-qubit rotation gates.

With our quantum resource, the variational quantum circuit in Eq.~(\ref{ansatz}) is set as 
$\hat{U}_\mathbf{\theta}=\prod_{k=1}^{d} \hat{U}_g \times \hat{u}(\theta^k_1) \times ... \times  \hat{u}(\theta^k_N)$.
 It contains $d=5$ layers of unitary operations, each of which is composed of a global entanglement gate $\hat{U}_g=\exp(-i\hat{H}_0\tau)$, where $\hat{H}_0=\sum_i^{N-1} g_{i,  i+1}(\hat{\sigma}^{+}_i \hat{\sigma}^{-}_{i+1} + \hat{\sigma}^{-}_i \hat{\sigma}^{+}_{i+1})$ and $\tau = 9$~ns, followed by variational single qubit $R_z$ gates $\hat{u}(\theta^k_i)$ for preserving the $U(1)$ symmetry. 
The circuit structures are shown in Fig.~\ref{fig:fig1}a, and the pulse sequence of the circuit is shown in Fig.~\ref{fig:fig1}c.

For the classical ansatz $p_{\phi}(\x)$, there are many choices such as a factorized distribution~\cite{DLbook}, restricted Boltzmann machine~\cite{Ackley1985,hinton2006reducing}, or the variational autoregressive model~\cite{frey1998graphical,pixelcnn,wu2019solving,liu_solving_2021}, building upon recent progress of unsupervised modeling in machine learning~\cite{DLbook}. As a proof of principle, we take 
a simple product Bernoulli distribution $p_\phi(\x)=\prod_i p_{\phi_i}(x_i)$  in this work (see Appendix~\ref{app_A} for more details), which is enough for demonstrating the scalable feature in computing thermal observables with a small statistical error.

 \begin{figure}[t]
	\includegraphics[width=0.5\textwidth=]{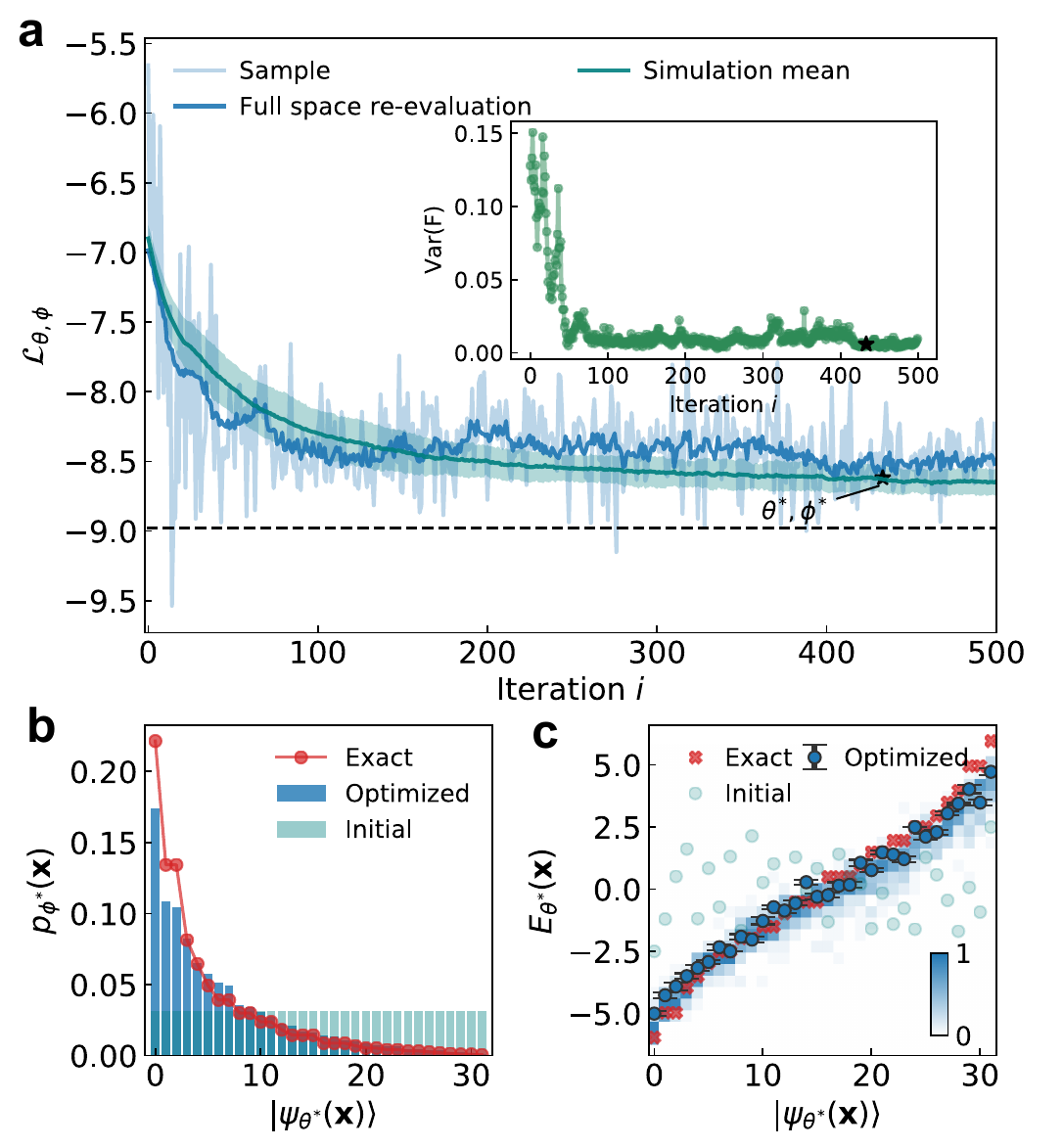}
	\caption{Experimental results for the quantum XY model with $h=0.5$, $\beta=0.5$. \textbf{a}, the evolution of the loss function $L_{\theta,\phi}$ evaluated using samples (Sample) and by enumerating all $2^5$ basis vectors (Full space re-evaluation) during the parameter learning. 
	The shaded area denotes the standard error of $50$ real-parameter numerical simulation curves using the full-space re-evaluated method . The black dashed line is the exact value of free energy. The inset shows that the sample variance of the loss function decreases during the parameters learning. The optimized distribution $p_{\phi^*}(\x)$ (\textbf{b}) and energy $E_{\theta^*}(\x)$ (\textbf{c}) after the parameter learning are illustrated and compared with the exact values.
	The energy values for $2^5$ basis vectors $\{\x\}$ are sorted by their probabilities in $p_{\phi^*}(\x)$. The density plot in \textbf{c} represents the eigenenergies obtained from 50 numerical simulation results. }
	\label{fig:fig2}
\end{figure}

\begin{figure*}[!tp]
	\includegraphics[width=1.0\textwidth=]{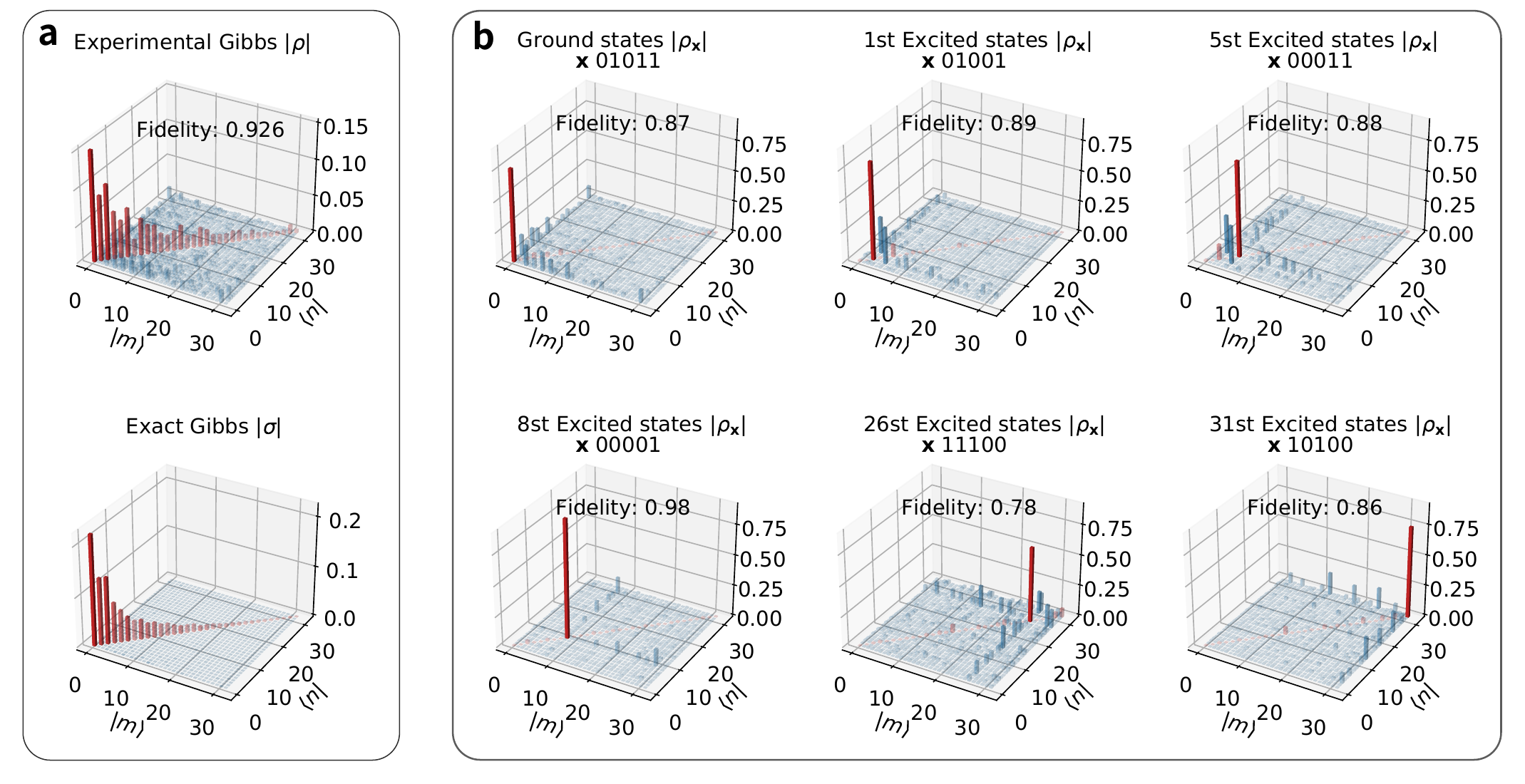}
	\caption{\textbf{Density matrix of the Gibbs states and eigenstates obtained in the experiments for the quantum XY model}. \textbf{a}, density matrix of the obtained Gibbs states $\hat{\rho}^*$ (upper panel) compared with the exact Gibbs states (lower panel), the fidelity is 92.6\%. \textbf{b}, density matrix $\rho_{\x}$ of experimentally prepared eigenstates $|\psi_{\theta^*}(\x)\rangle$, where the corresponding exact eigenstate $|n\rangle$ is identified by the $n$th probabilities $p_{\phi^*}(\x)$ sorted in a descent order. In all the plots, only the amplitudes of the density matrix are shown, with diagonal elements and off diagonal elements colored with red and blue respectively. 
	}
	
	\label{fig:fig3}
\end{figure*}

\section{Results}
Here we present the experimental results for quantum XY chain defined in Eq.~(\ref{Hamil}) with $\Delta=0$ and $h=0.5$. 
In our approach, the sample size can be 
extremely small when compared to the total Hilbert space dimension. For a 5-spin chain in the experiments, 
we could take only $2$ samples from $p_\phi(\x)$ in each learning step as an input product states for the quantum circuit $U_\theta$, and for computing the loss function and its gradients.
The number of samples is much smaller than the dimension of the Hilbert space which is $2^5$, thus indicating a possible exponential reduction of computational complexity. We confirm this by a numerical study of the computational complexity and show that the total time cost grows polynomial in qubit number (see Appendix~\ref{app_E}).
It turns out that such a small sample size already gives an accurate result and a small variance of the loss function.
As a sanity check, we have also computed the loss function using the parameters learned with the sampling approach, by enumerating all $2^5$ basis vectors in the computational space, which we term as \textit{full-space re-evaluation}, for evaluating the performance of the learning procedure using samples. 

\textit{Optimization results and self-verification --} We present the optimization trajectory at $\beta=0.5$ in Fig.~\ref{fig:fig2}a. The loss function given in both experiments and real-parameter numerical simulations decreases from a large initial value to a value that is about 5\% higher than the exact free energy at the end of learning. Based on our experimental error model (see Appendix~\ref{app_C}), the experimental results agree with the numerical simulation results well.
The final parameters of the $U_\theta$ and $p_\phi$ are determined using the $\theta^*, \phi^*$ values corresponding to the minimal loss in the full-space re-evaluation curve. 
In Fig.~\ref{fig:fig2} (b) and (c) we present the probability distribution $p_{\phi^*}(\x)$ and energy expectation $E_{\theta^*}(\x)$ for each participate states of the optimized Gibbs ensemble, which are in coincidence well with the exact distribution $P_{\text{Gibbs}}(n)=e^{-\beta E_n}/\mathcal{Z}$ and the eigen energies $E_n$.
For comparison, we also perform an ideal noiseless simulation with exact quantum gates. The results are shown Appendix ~\ref{app_H} Fig.~\ref{fig:ext_fig2}, where the probability distribution and the eigen energies match the exact values very closely, indicating that our quantum circuit can fully diagonalize the target Hamiltonian in the noiseless case. 

We now demonstrate that the self-verification of the optimization can be done by the sample variance. It is shown both in Fig.~\ref{fig:fig2}a and Appendix ~\ref{app_H} Fig.~\ref{fig:ext_fig2} that the variance of the sample curve is larger than that of the full-space curve and decreasing during the optimization (see the inset of Fig.~\ref{fig:fig2}a). This is because that when the learning is exact, the variance should be zero.
 Thus empirically we can use the sampling variance of the loss function as a self-verifying indicator on the quality of parameter learning. Moreover, this indicator can be accessed directly by sample method hence avoid extra complex measurement that needed in the ground state VQE algorithm~\cite{kokail_self-verifying_2019}. In Appendix~\ref{app_F}, we give a more detailed analysis of the self-verification using the variance argument.

\textit{Preparation of Gibbs states and excited states --} The learned variational ansatz allows us to prepare the Gibbs states of $\hat{\rho}^*=\sum_x p_{\phi^*}(\x) |\psi_{\theta^*}(\x)\rangle \langle\psi_{\theta^*}(\x)|$ by mixing the component states $|\psi_{\theta^*}(\x)\rangle$ and their probabilities $p_{\phi^*}(\x)$. 
By state tomography, we obtained the variational Gibbs states with fidelity 92.6\% to the exact Gibbs state. The obtained density matrix of the Gibbs state in the eigenbasis $\{|n\rangle\}$ of target Hamiltonian is shown in Fig.~\ref{fig:fig3}a. Furthermore, since in our approach the probability of each state can be determined, we can hence distinguish  excited states (up to degenerations) using their probability values. This allows us to experimentally prepare both the ground state and specified excited states of the target Hamiltonian, with the optimized diagonalization unitary $\hat{U}(\theta^*)$. More specifically, the exact eigenstate $|n\rangle$ corresponding to $|\psi_{\theta^*}(\x)\rangle$ is identified using the bitstring with $n$-th largest probability in $p_{\phi^*}(\x)$. 
In Fig.~\ref{fig:fig3}b,  the density matrix $\hat{\rho}_\x$ of several experimentally prepared excited states in the eigenbasis $\{|n\rangle\}$, together with the corresponding initial basis $\x$ are presented. We observe that each density matrix has the largest amplitude located at the correct entry in the density matrix, with high fidelity to the corresponding exact density matrices.
The fidelity matrix between the obtained eigenstates $|\psi_{\theta^{*}}(\mathbf{x})\rangle$ and the exact eigenstates $|n\rangle$ are illustrated in Appendix ~\ref{app_H} Fig.~\ref{fig:ext_fig3}.

\begin{figure}[!tp]
	\includegraphics[width=0.5\textwidth=]{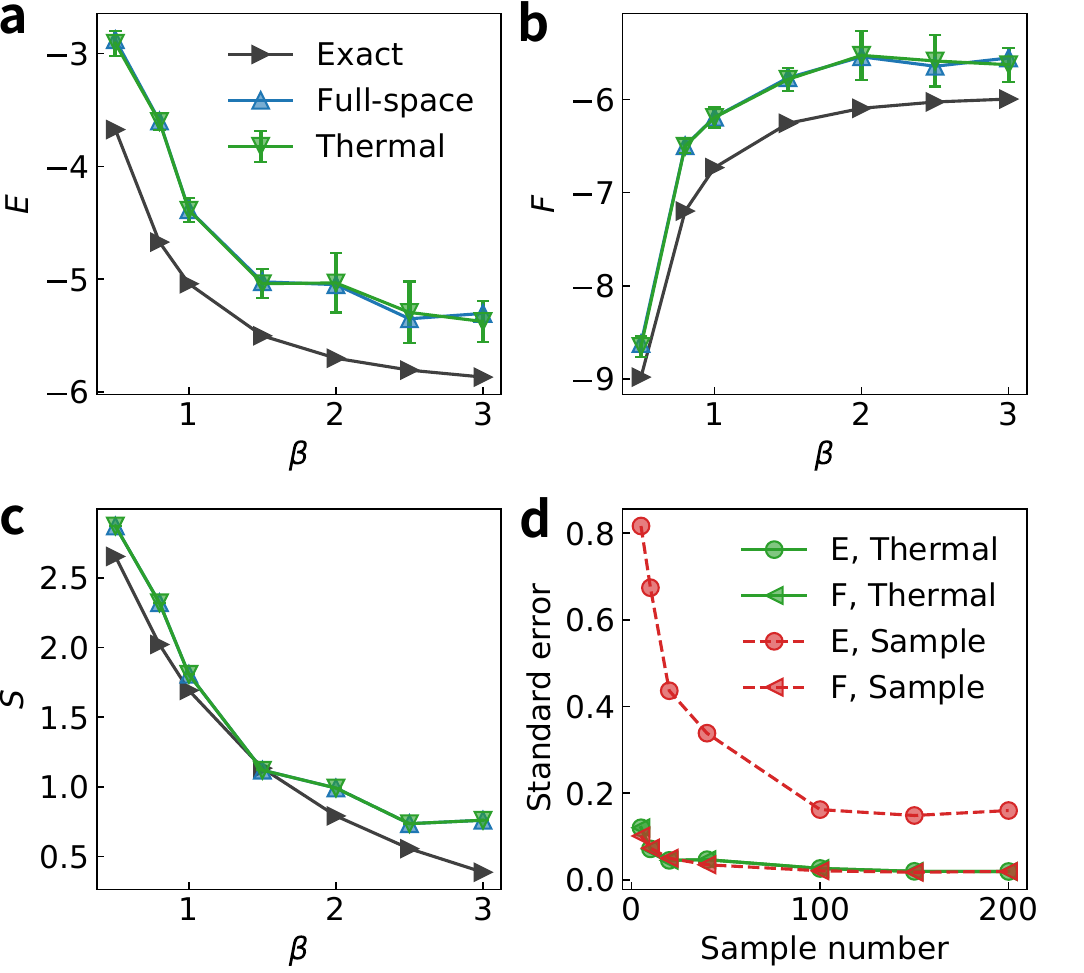}
	\caption{\textbf{Thermal quantities obtained by experiment at different inverse temperature}. \textbf{a}, \textbf{b}, the internal energy (\textbf{a}), free energy (\textbf{b}) derived by three different method: full-space (up triangle) , sample averaged (circle) and thermal relation approach (down triangle). In the last two methods, the sample size is 5 and the error bars is obtained as the standard error by repeating the evaluation for 20 times. 
	The entropy (\textbf{c}) is evaluated using an analytical expression hence contains no statistical error. \textbf{d}, standard error of experimental derived internal energy and free energy as a function of sample number. Statistical error of internal energy with sample averaged method is much larger than the thermal approach.
	}
	
	\label{fig:fig4}
\end{figure}

\textit{Thermal observables estimation --} In addition to the free energy, thermal quantities such as the energy and the entropy can also be estimated using samples according to $E = \mathbb{E}_{x \sim p_{\phi} (\x)} [\langle \psi_{\mathbf{\theta}}(\x)|\hat{H} |\psi_{\mathbf{\theta}}(\x)\rangle]$ and $S = \mathbb{E}_{x \sim p_{\phi} (\x)} [\ln p_{\phi}]$ respectively. 
These observables generally have larger statistical variance than the free energy $F$ and cannot be estimated precisely with a small number of samples. For the energy estimation
, if we still use a sample number that much smaller than the Hilbert space dimension as in the optimization process, the sample variance and statistical error would be quite large, as illustrated in Fig.~\ref{fig:fig4}d. The same situation is also encountered in other approaches e.g. in~\cite{sun_quantum_2021}. 
Here we propose to resolve this issue by utilizing the thermodynamic relation $E=F+\frac{1}{\beta}S$ , based on the self-verifying feature of $F$ and the classical estimation of entropy S (see Appendix~\ref{app_B}). In this way, the variance of the energy estimation can be greatly reduced. In Fig.~\ref{fig:fig4}a, b, c, we plot the thermal quantities obtained in experiments where the entropy is computed using an analytical method with no statistical error and the free energy is evaluated using $5$ samples. The variance of energy is at the same level as those of free energy and entropy, which are much smaller than those obtained in existing methods. From Fig.~\ref{fig:fig4}d, the standard error for internal energy obtained by the sample method with 200 samples is even larger than that obtained by our improved approach with 5 samples. Thus, our strategy is efficient for large-scale problems and is extendable to other thermal quantities which can be expressed as a function of free energy and entropy.

Moreover, Fig.~\ref{fig:fig4} also shows that our approach can successfully prepare Gibbs states and obtain thermal quantities in a wide range of temperatures, particularly when compared with existing approaches. For example, the truncated entropy approach~\cite{wang_variational_2020} is limited to low-temperature regions due to the series truncation of entropy, and the quantum imaginary time evolution method~\cite{motta_determining_2020, sun_quantum_2021} suffers from error accumulation at a low temperature due to the fast-growing circuit depth resulted by a long imaginary evolution time, hence has difficulties in reaching an accurate estimate of energy as long as $\beta\geq 0.5$. In contrast, our algorithm uses a quantum circuit with a fixed depth, hence is not influenced significantly by temperature and works all the way down to $\beta=3.0$ as shown in Fig.~\ref{fig:fig4}.

\textit{Results for XXZ model. --} Finally, to further demonstrate the performance of our variational approach, we apply the same method to the spin-1/2 XXZ chain with $h=0$, which is an interaction model and has an additional $\mathbb{Z}_2$ symmetry. We find that our $U(1)$ preserving quantum circuit also produces reliable results for the XXZ chain with the model parameters we studied (see Appendix ~\ref{app_H} Fig.~\ref{fig:ext_fig1}), indicating that our method is a general approach for studying different quantum lattice models.
\section{Discussions}
We have demonstrated the hybrid quantum-classical variation approach on a superconducting quantum processor for solving quantum statistical mechanics of the quantum XY and XXZ chains.
The method utilizes the generative probabilistic modeling in machine learning for maintaining the mixture distribution and estimating the variational entropy, and the quantum processor for performing unitary transformations and estimating the energy, thus takes advantage of both classical and quantum resources. The parameters of the generative model and quantum circuits are learned through a mini-batch gradient-based method.
We have shown that the variational approach can prepare Gibbs states and excited states for the XY and XXZ models with high fidelity, with a self-verifiable feature using the variance of the loss function, and can estimate thermal quantities with a small statistical error.

Our approach is general and flexible for extensions. The quantum circuits can be readily updated to near-term quantum devices with a much larger size, and the classical distribution can be generalized to more representative neural network generative models straightforwardly~\cite{wu2019solving,liu_solving_2021}.
On the application side, the proposed approach can be extended immediately to other condensed matter models such as the Fermion systems using the Jordan-Wigner transformation.  
Moreover, with the prepared Gibbs states, we can further investigate the finite-temperature dynamics on quantum simulators~\cite{sun_quantum_2021}. Last but not least, the preparation of eigenstates at certain energy densities also makes it possible to study many-body localization~\cite{spall_implementation_1998, abanin_colloquium:_2019, guo_observation_2021-1} and eigenstates thermalization hypothesis~\cite{deutsch_quantum_1991, srednicki_chaos_1994} with quantum computers. 

\vspace{6mm}
\noindent\textbf{{Acknowledgements}} 
We thank Zhengan Wang, Ruizhen Huang and Tao Xiang for useful discussions. This work is supported by National Science Foundation of China (Grant Nos. 11934018, 11747601, 11975294, 92065114, 11904393), Strategic Priority Research Program of Chinese Academy of Sciences (Grant No. XDB28000000), Scientific Instrument Developing Project of Chinese Academy of Sciences (Grant No. YJKYYQ20200041), Beijing Natural Science Foundation (Grant No. Z200009), and CAS Grant No. QYZDB-SSW-SYS032.

\vspace{6mm}
\noindent\textbf{{Data availability}}
All data generated and analyzed in this study are available on \href{https://github.com/xiaoxiao9689/TVQS}{https://github.com/xiaoxiao9689/TVQS}.

\vspace{6mm}
\noindent\textbf{{Code availability}}
The optimization algorithm developed during this study and simulation code is available \href{https://github.com/xiaoxiao9689/TVQS}{https://github.com/xiaoxiao9689/TVQS}.

\appendix

\section{Classical optimization scheme.}\label{app_A}
As mentioned in the main text, we sample from a variational distribution and using a few samples in computing the loss function and observables, rather than considering the total $2^N$ computational basis vectors.
Due to the large fluctuations of the loss function estimated using a small number of samples, the gradient-free optimizers such as the Nelder-Mead simplex method, particle swarm algorithm, Bayesian optimization~\cite{zhu_training_2019}, and the dividing rectangles optimizer~\cite{kokail_self-verifying_2019} are not suitable for our task. In this work, we consider the gradient-based optimization scheme, with gradients computed as

\begin{align}
	\nabla_{\theta} \mathcal{L} & =  \mathbb{E}_{x \sim p_{\phi}} [
	\nabla_{\theta}\langle \x|\hat{U}_{\theta}^{\dag} H \hat{U}_{\theta} |\x \rangle] \label{grad1}, \\
	\nabla_{\phi} \mathcal{L} & =  \nabla_{\phi} \mathbb{E}_{x \sim p_{\phi}} \left[\frac{1}{\beta}\ln p_{\phi} (\x)+\langle \x|\hat{U}_{\theta}^{\dag} H \hat{U}_{\theta} |\x \rangle\right], 
	\label{grad2}
  \end{align} 
In general, one cannot compute directly the gradients with respect to the model parameters, so we need to use the score function gradient estimator
\begin{align}
	\nabla_{\phi} \mathcal{L}  =  \mathbb{E}_{x \sim p_{\phi}} [(R(\x) - b) \nabla_{\phi} \ln p_{\phi} (\x)], \label{grad3}
\end{align}
where $R(\x) = \ln p_{\phi} (\x)/{\beta} + 
	\langle \psi_{\mathbf{\theta}}(\x)|\hat{H} |\psi_{\mathbf{\theta}}(\x)\rangle$, $b$ denotes a \textit{base line} parameter to reduce the variance~\cite{wu2019solving,liu_solving_2021,williams1992simple,DLbook,Mnih2014a}, and we adopt a common choice that $b = \mathbb E_{x\sim \phi}\mathcal R(\x)$.
To understand the formulation, in computing the expectations, each sample $\x\sim p_\phi(\x)$ is weighted by $R(\x)$, the (negative) reward function, in the sense that the gradients would be large and the optimization would try to reduce the probability of generating the sample $\x$ if the corresponding (negative) rewards are large.
The algorithm is also known as the REINFORCE algorithm in machine learning~\cite{williams1992simple,sutton1998reinforcement}.

To estimate the gradients of the energy $\nabla_{\theta}\langle \x|\hat{U}_{\theta}^{\dag} H \hat{U}_{\theta} |\x \rangle]$ with respect to parameters of the quantum circuit $\theta$ (which are the parameters of the single-qubit rotation gates in our set up), we could employ the parameter shift rule (PSR)~\cite{schuld_evaluating_2019}, which only requests the loss function values at a forward and backward shift of $\pi/2$ for each parameter. However, it is inefficient for a quantum circuit containing a large number of parameterized gates and has limitations when the variational gates do not satisfy the PSR condition.
Another popular method to estimate the gradients is the simultaneous perturbation stochastic approximation (SPSA) algorithm~\cite{spall_adaptive_2000}. It approximates the gradients of the loss function for all parameters simultaneously by taking only two values of the loss function in a perturbation way, hence has a constant time complexity regardless of the number of gates or shape of quantum circuits.
However, in the experiments, we observed that the computational cost of pure SPSA methods depends on the number of iteration steps $n_{\text{iter}}$ to achieve the desired accuracy, which follows a power-law scaling and has a high chance of trapping into local minima of the loss function.

In this work, we use a hybrid gradient descent optimization scheme by combining the SPSA with the adaptive moment estimation (Adam)~\cite{adam} optimizer, denoted by SPSA-Adam. In the algorithm, the exact gradients in the Adam optimizer are replaced by the mean of several SPSA gradient estimations~\cite{spall_adaptive_2000}. The hybrid optimizer combines together the advantages from the compatibility of the SPSA gradients and the fast convergence in the noisy environment from the Adam optimizer.  
In the experiments, we observed that the number of SPSA trails (for obtaining the average gradients) $n_\text{SPSA}$ has a polynomial scaling  to the number of qubits $N$ (Appendix~\ref{app_E}).

For the classical distribution, in consideration that the number of qubits is small, we use a product of Bernoulli distribution for each qubit
\begin{equation}
	p_{\phi} (\x) = \prod_i p_{\phi_i} (x_i) = \prod_i \phi_i^{x_i} (1 - \phi_i)^{1 - x_i},
\end{equation}
where $\phi_i \in [0, 1]$ is the probability that $x_i$ being $1$. 
Each Bernoulli distribution is further parameterized using the sigmoid function with a single variable $\phi_i(x) = 1/(1+e^{-x})$. 
The gradients for $\phi$ are then derived ``semi-classically", where the evaluation of $\nabla_{\phi} \ln p_{\phi} (\x)$ is done on classical computer where $R(\x)$ and $b$ contain the energy values measured using the quantum circuit.  
In the experiments, we observed that the product distribution already has enough expressive power for the system with $5$ qubits.
It is straightforward to replace the product distribution with a more powerful probabilistic model, e.g. the autoregressive model as adopted in Ref.~\cite{liu_solving_2021} which has much more parameters and a better representation power than the product distribution, by taking advantage of the deep learning techniques.

\section{Thermal observable calculations} \label{app_B}
After learning, the thermal observables such as the internal energy $E$ can be computed as a function of $p_\phi$ and $U_\theta$, using a large number of samples than those in the learning process. The self-verifying feature implies that when the obtained variational free energy is close to the exact values, the small variance of the samples allows us to accurately estimate $F$ using a small number of samples.
Remarkably, since entropy estimation only relies on the classical distribution using a generative model, observables such as the entropy which are hard to compute on quantum computers can be estimated efficiently classically using $p_\phi(\x)$.
In particular, benefit from the product distribution ansatz of $p_\phi(\x)=\prod_i p_{\phi_i}(x_i)$, we are able to evaluate the entropy with no statistical error using an analytical expression 
\begin{align}
S=-\sum_i\sum_{x_i}p_{\phi_i}(x_i)\log(p_{\phi_i}(x_i)).
\end{align}
If the classical distribution is implemented using a more representative model such as the autoregressive neural networks~\cite{liu_solving_2021}, one can still use a polynomial algorithm to generate a large number of samples for estimating the entropy, resulting in a much smaller statistical error than that of energy which requires access to a quantum resource (see Appendix~\ref{app_G}). Thus, given the thermal relation $F=E-\frac{1}{\beta}S$, we can compute $E$ more accurately with a much smaller statistical error than estimating it directly using samples.

\section{The error model for the quantum device.}\label{app_C}
 The Hamiltonian used in Eq.~\eqref{expH} is an approximate version of the real quantum device Hamiltonian. There are two intrinsic reasons that make the simulations on the hardware deviate from the ideal simulation results. One reason is the extra small couplings between the next-nearest neighbor qubits. Since our entanglement layer is realized in an analog way, the numerical simulation results show that in the presence of next-nearest neighbor couplings the ansatz has a poorer expression of the target model. The problem can be resolved by using two-qubit gates to realize entanglements. Another reason is the state-leakage error from non-infinite anharmonicity of the transmon qubit, where our device should be modeled by the Bose-Hubbard model with a large on-site interaction. 
The  state-leakage error not only increases the measurement error of the observables but also prevents us from accessing regions with a very high temperature. The issue comes from the requirements of unitarity of the quantum circuit $\hat{U}_{\theta}$ for evaluating the entropy part in loss function classically. To extract the spin observable with a  measured $3^5$ dimensional density matrix, we abandon the probabilities that one or more qubits are excited to the $|2\rangle$ states, leading to an effective non-unitary circuit. The non-unitary effect causes inaccuracy in the classical computation of entropy and destroys the variational principle. At high temperatures, the effect is more serious since the entropy plays a more important role in the loss function. Thus we present the results with $\beta\geq 0.5$, where the optimizations can proceed effectively.

There are mainly several possible resources contributing to the statistical errors in the measurements of observables. The first one is the measurement error when using many single-shot measurement results to calculate observables. The second one is the experimental random noise that leads to parameters fluctuation such as the idle frequency drift. The first error can be efficiently simulated while the latter is hard. But they all effectively generate random noise in observable evaluations. So in the simulations, we use smaller single-shot measurements to effectively simulate these two errors.

Another error resource is the decoherence process. Our circuit is shallow, the total operation time for the quantum gate is much less than the decoherence time $T_1$ and $T_2$ (see next section), so we neglect the decoherence effect in the real-parameter simulations.
\begin{table}[ht]
	\centering
	\scriptsize
	\renewcommand\arraystretch{1.6}
	\resizebox{0.4\textwidth}{!}{
		\begin{tabular}{c |c c c c c }
			\hline
			\hline
			
			~ &Q$_1$  &Q$_2$  &Q$_3$  &Q$_4$  &Q$_5$\\
			\hline
			$\omega_i^0/2\pi$~(GHz) &5.531  &4.968  &5.433  & 4.999 &5.502\\			
			$\omega_i/2\pi$~(GHz)  &5.435  &4.932  &5.378  &4.975  &5.471\\			
			U/2$\pi$~(MHz)  &-242  &-196  &-239  &-196  &-242\\
			T$_{1,i}$~($\mu$s)  &31  &35  &35  &36  &54\\
			T$_{2,i}^*$~($\mu$s)  &9.14  &7.39  &7.27  &8.74  &$12.64$\\
			\hline
			g$_{1,2}/2\pi$~(MHz) &\multicolumn{2}{c}{14.60} &~ &~  &~    \\
			g$_{2,3}/2\pi$~(MHz) &~& \multicolumn{2}{c}{14.65}  &~  &  ~    \\
			g$_{3,4}/2\pi$~(MHz) &~&~& \multicolumn{2}{c}{14.17}  &~     \\
			g$_{4,5}/2\pi$~(MHz) &~&~&~& \multicolumn{2}{c}{14.26}    \\
			\hline
			g$_{1,3}/2\pi$~(MHz) &\multicolumn{3}{c}{1.142}  &~  &~    \\
			g$_{2,4}/2\pi$~(MHz) &~& \multicolumn{3}{c}{0.607}    &~    \\
			g$_{3,5}/2\pi$~(MHz) &~&~& \multicolumn{3}{c}{1.207}    \\
			\hline
			$\omega^r_i/2\pi$~(GHz)  &6.612  &6.654  &6.687  & 6.7266 &6.766\\
			F$_{0,i}$  &0.98  &0.97  &0.97  &0.97  &0.99\\
			F$_{1,i}$  &0.91  &0.88  &0.89  &0.89  &0.89\\
			
		\end{tabular}
	}
	\caption{Qubit characteristics.
		$\omega_{\textrm{i}}^{\textrm{0}}$ is the zero flux biased frequency of Q$_i$. $\omega_{\textrm{i}}$ is the idle
		frequency of Q$_i$. $\omega^{\textrm{r}}_i$ is the readout resonator frequency of Q$_i$. T$_{1,i}$ is the energy relaxation time of Q$_i$ at the idle frequency. T$_{2,i}^*$ is the dephasing time of Q$_i$ at the idle frequency. U/2$\pi$ is the non linearity ($f_{21}-f_{10}$) of Q$_i$ measured at the zero flux bias. F$_{1,i}$ (F$_{0,i}$) is the measured probability of $|1\rangle$ ($|0\rangle$) when $Q_i$ is prepared in $|1\rangle$ ($|0\rangle$). g$_{i,j}$ is the coupling strength between Q$_i$ and Q$_j$.}
	\label{tab1}	
\end{table}
\section{Experimental details.} \label{app_D}
\noindent\emph{(1) Experiment setup.} Our quantum device is placed at 10mK in a BlueFors Dilution Refrigerator. It is well screened from a higher temperature environment with its own aluminum alloy package box and a magnetic shield. Outside of which, we have another layer of aluminum alloy shield and magnetic shield at 10mK. All these shields are tightly sealed and well thermally connected to the 10mK plate. The control wires connected to the quantum device are deeply attenuated and filtered against noise from the room temperature environment and electronic setups. For XY control lines~\cite{barends_coherent_2013}, we have 3dB attenuation at 40K, 20dB attenuation at 3K, 6dB attenuation at 800mK, and 40dB attenuation at 10mK, plus a 7.5GHz low pass filter. For Z control lines~\cite{barends_coherent_2013}, we use a capacitance removed Bias-Tee at 10mK to combine the DC bias and fast Z square pulse. For the fast Z control lines, we have 3dB attenuation at 40K, 20dB attenuation at 3K, 6dB attenuation at 800mK, and 10dB attenuation at 10mK, plus a 500MHz low pass filter before connected to the Bias-Tee. We use continuous 0.86mm CuNi coaxial cables as our DC bias lines, and apply an RC low-pass filter at room temperature, where R = 1K$\Omega$, and a 200MHz low-pass RLC filter at 10mK before connected to the Bias-Tee. For the input line of readout pulses, we have 3dB attenuation at 40K, 20dB attenuation at 3K, 6dB attenuation at 800mK, and 40dB attenuation plus a 7.5GHz low-pass filter at 10mK. For the output line of readout pulses, we have a Josephson parametric amplifier (JPA) at 10mK, a HEMT amplifier at 3K, and another microwave amplifier at  room temperature. We isolate our quantum device from the JPA with a 7.5GHz low-pass filter and an isolator. We have another isolator between the JPA and HEMT.  
\begin{figure}[t]
	\centering
	\includegraphics[width=7.5cm]{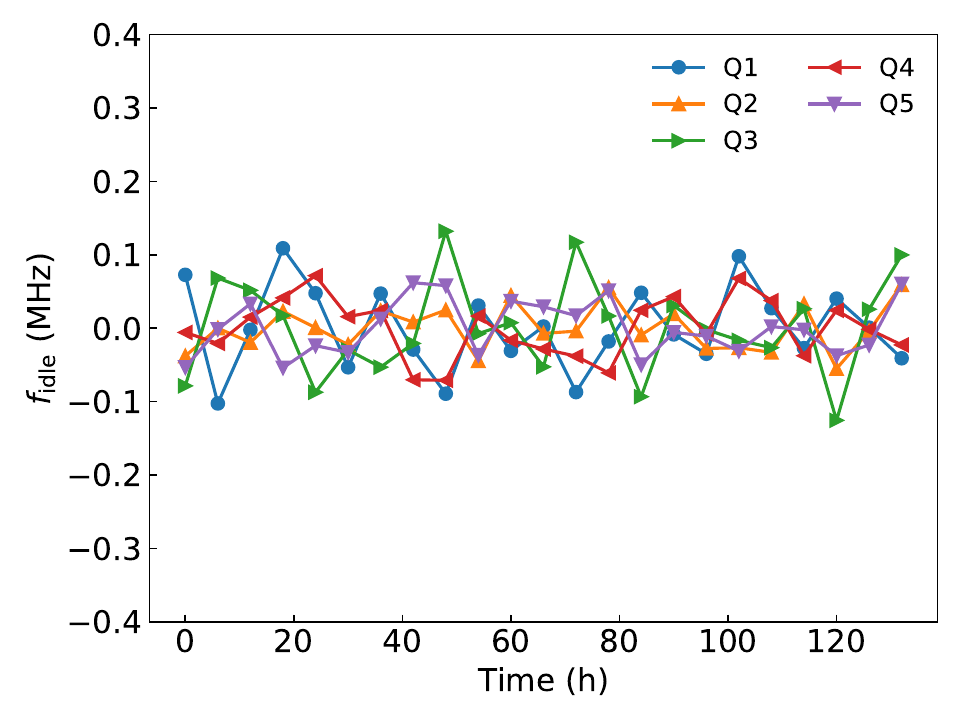}
	\caption{Fluctuation of the qubit idle frequencies. We have monitored the idle frequency of all qubits and found that the fluctuations are in the range [-0.1,0.1] MHz, which is close to the adjusted precision of our room temperature electronic system.}
	
	\label{fig:ext_fig6}
\end{figure}
Room-temperature electronic instruments are used to generate stable direct current, microwave pulses, and current pulses to control and readout states of qubits. Here, we use Yokogawa GS220 as a DC voltage source to bias qubits to their idle frequencies, the output range is set to 1V.  Zurich instrument HDAWGs are used to generate microwave pulses via IQMixers, and to generate current pulses as fast Z control pulses.

Our quantum device is shown in Fig.\ref{fig:fig1}a. and basic characteristics of qubits are listed in Appendix Table~\ref{tab1}.
 Compared to the previous experiment~\cite{guo_observation_2021}, the idle frequency $\omega_i$ of each qubit is closer to its sweet point $\omega_i^0$. With this idle frequency setup, the dephasing times $T_{2}^*$ at idle points of 5 qubits can reach 7-12 $\mu s$. The average energy relaxation time $T_{1}$ at idle points of five qubits is 38.2 $\micro s$.
While the resonant coupling frequency in this experiment is 4.932GHz. 
To tune all five qubits from their idle frequency to 4.932GHz, the output range for Z control channels of our Zurich instrument HD is set to 2V instead of 800mV from its direct mode. We found that this change brings little affection to the dephasing time of the qubits.     

Generally speaking, the variational optimization process is robust to coherent errors while still be affected by random interference and noise from the environment. With this experimental setup, we monitored the fluctuation of idle frequency for days during the experiment and found that the fluctuation of all 5 qubits is in the range of $[-0.1,0.1]$ MHz, as shown in  Fig.\ref{fig:ext_fig6}. This fluctuation amplitude is close to the adjustable precision of our experiment setup \cite{guo_observation_2021}.

\begin{figure}[t]
	\centering
	\includegraphics[width=7.5cm]{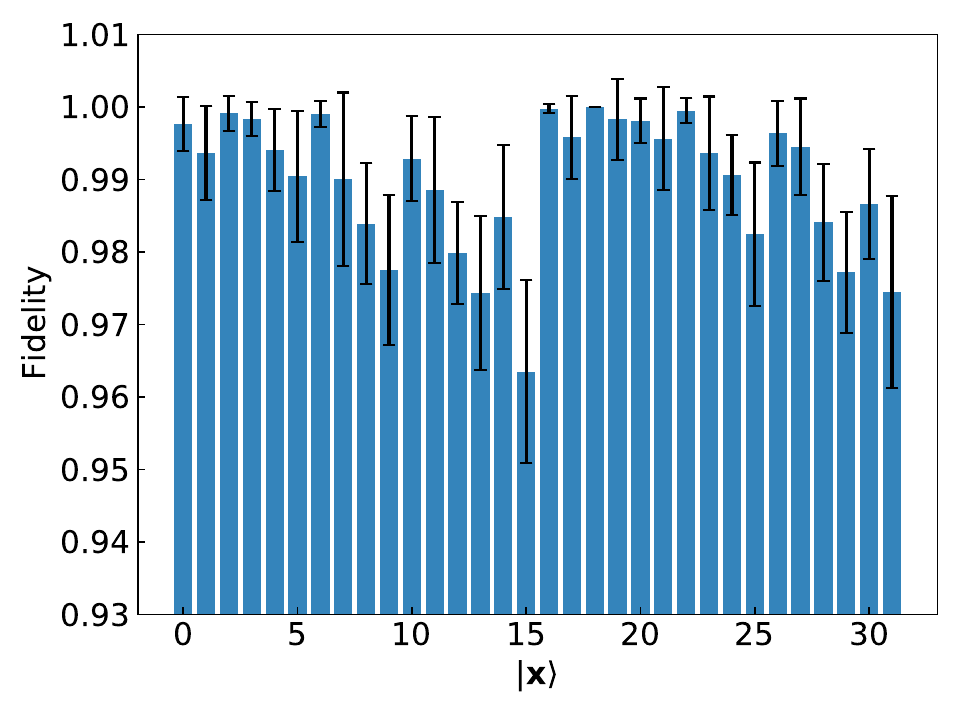}
	\caption{Fidelity of the input states. The $2^5$ computational basis states are prepared using a combination of $X$ gates and the fidelity of the states are measured. Preparing multi qubits in $|1\rangle$ suffers from the unwanted crosstalk affection induced from multi $X$ gates and the residual coupling between qubits which reduce the fidelity of the prepared states.}
	
	\label{fig:ext_fig7}
\end{figure}

\noindent\emph{(2) Preparation of computational basis states.} In the experiments, the input states for the quantum circuits are chosen from the set of computational basis $\{|00000\rangle,\cdots,|11111\rangle\}$, which are generated with combinations of $X_i$ gates.  We use the optimization methods in reference \cite{lucero_reduced_2010} to reducing the phase error of $X$, $X/2$, and $Y/2$ gates used in our experiment. Preparing multi qubits in $|1\rangle$ suffers from the unwanted crosstalk affection induced by multi $X$ gates, and the residual coupling between qubits. These effects reduce the fidelity of the prepared states. The measured fidelity of all $2^5$ input states is presented in  Fig.\ref{fig:ext_fig7}.  

\noindent\emph{(3) Constructing parameterized quantum circuit.}  The variational quantum circuit is constructed by layers of global entanglement gate and single-qubit $R_z$ rotation gates. To realize the entanglement gate, fast square pulses are applied to tune all qubits to the same frequency for $\approx 9$ ns, corresponding to dimensionless time $\tau=\pi/4$.  With the measured step response function of each fast Z control line, we can compensate and correct the distortion of applied square pulses, meanwhile, eliminate the fluctuations after the square pulses\cite{guo_observation_2021}. 
The programable single-qubit gates in the experiments are realized using the $R_z$ gates, by tuning qubit frequency with amplitude tunable square pulses of duration 25 ns. The R$_z$ gates are calibrated using the Ramsey-like experiments. We insert an $R_z$ gate between two $\frac{\pi}{2}$ pulses. For a given amplitude of the $R_z$ gate, the phase of second $\frac{\pi}{2}$ pulse is verified to determine the rotation angle, which results in a function of the rotation angle and pulse amplitude, using which we can set the required rotation angle of R$_z(\theta)$ in our experiments. 
The crosstalk between the Z control lines are experimentally determined and compensated so that the R$_z$ gates for different qubits have no affection to each other~\cite{guo_observation_2021}. 

\noindent\emph{(4) Energy expectation determination.} 	The energy expectation $\langle H \rangle$ of the objective Hamiltonian $H$ is determined by measuring and summing all terms $\langle \sigma_x^i\sigma_x^{i+1} \rangle$, $\langle \sigma_y^i\sigma_y^{i+1} \rangle$, $\langle \sigma_z^i\sigma_z^{i+1} \rangle$, $\langle \sigma_z^i \rangle$.  For the $\langle \sigma_x^i\sigma_x^{i+1} \rangle$  and $\langle \sigma_y^i\sigma_y^{i+1} \rangle$ terms, single-qubit rotations $X/2$ ($Y/2$) are performed before readout. 
In the experiments,  we measure all qubits simultaneously  by rotating all qubits before a joint readout. Thus, in each measurement of energy expectation we only need to rerun the circuit for three times in order to compute $\{ \langle \sigma_x^i\sigma_x^{i+1} \rangle \}$, $\{ \langle \sigma_y^i\sigma_y^{i+1} \rangle \}$, and $\{ \langle \sigma_z^i\sigma_z^{i+1} \rangle, \langle \sigma_z^i \rangle \}$.

For the input states with more than one qubit in an excited state, the implementation of a global entanglement gate can induce the state leakage to $|2\rangle$, as described in the last section. To reduce the state-leakage error, we measure all qubits simultaneously under the qutrit computational basis. Thus, the dimension of measured Hilbert space is $3^5$, i.e. from $|00000\rangle$ to $|22222\rangle$. Then, we reduce and normalize the measured Hilbert space to $2^5$ under the qubit computational basis, by discarding the results of one or more qubits in $|2\rangle$.  It should be pointed out that such state-leakage error induced in resonant interaction-based entanglement gates could be effectively reduced by using non-resonant interaction-based entanglement gate schemes, such as conditional-phase gates based on tunable ZZ-interactions~\cite{herrmann_implementation_2020}.

\section{Numerical results on scalability}\label{app_E}
In this section, we present an analysis of the computational complexity of the variational algorithm in numerical simulations with qubit number $N$ ranging from 4 to 10. As mentioned in Appendix~\ref{app_A}, there are several classical optimization schemes depending on which gradients it uses, e.g. the sampling method or the SPSA gradients. The optimizers have different hyperparameters giving different performance and resource costs. We determine the hyperparameters and corresponding resource cost for a given precision starting from the algorithm containing a minimum number of hyperparameters. First, by using the full-space approach and the precise gradients derived by parameter shift, denoted as the full-space PSR-Adam optimizer, we determine the scaling of circuit depth $n_{\text{layer}}$ with respect to qubit number $N$. Next, by selecting the value of $n_{\text{layer}}$ on the contour line of a certain precision, we employ the sampling version of PSR-Adam to determine the scaling of cost with different sample sizes $n_{\text{batch}}$. 
Finally, with $n_{\text{layer}}$ and $n_{\text{batch}}$ determined in the above steps, the scaling behavior of sampling SPSA-Adam scheme is studied, where an additional hyperparameter $n_{\text{SPSA}}$, denoting the average number of SPSA gradient evaluation, is considered. The final cost of the optimization scheme shows a polynomial growth, which means that our approach is scalable for large systems. Although the results are obtained under a target model of the XY chain, the experimental and numerical results for the XXZ model (see Appendix ~\ref{app_H} Fig.~\ref{fig:ext_fig1}) confirm that the scaling is the property of the optimization algorithm rather than the target model, hence we expect that the scaling is universal for other complex models. 
Here we present details of the scaling analysis on the XY model where both $h$ and $\beta$ are set to be 0.5. To save computational resources in both time and space, we did not consider noises in the numerical simulation.

\subsection{Full-space PSR-Adam}
\begin{figure*}[t]
	\centering
	\includegraphics[width=14cm]{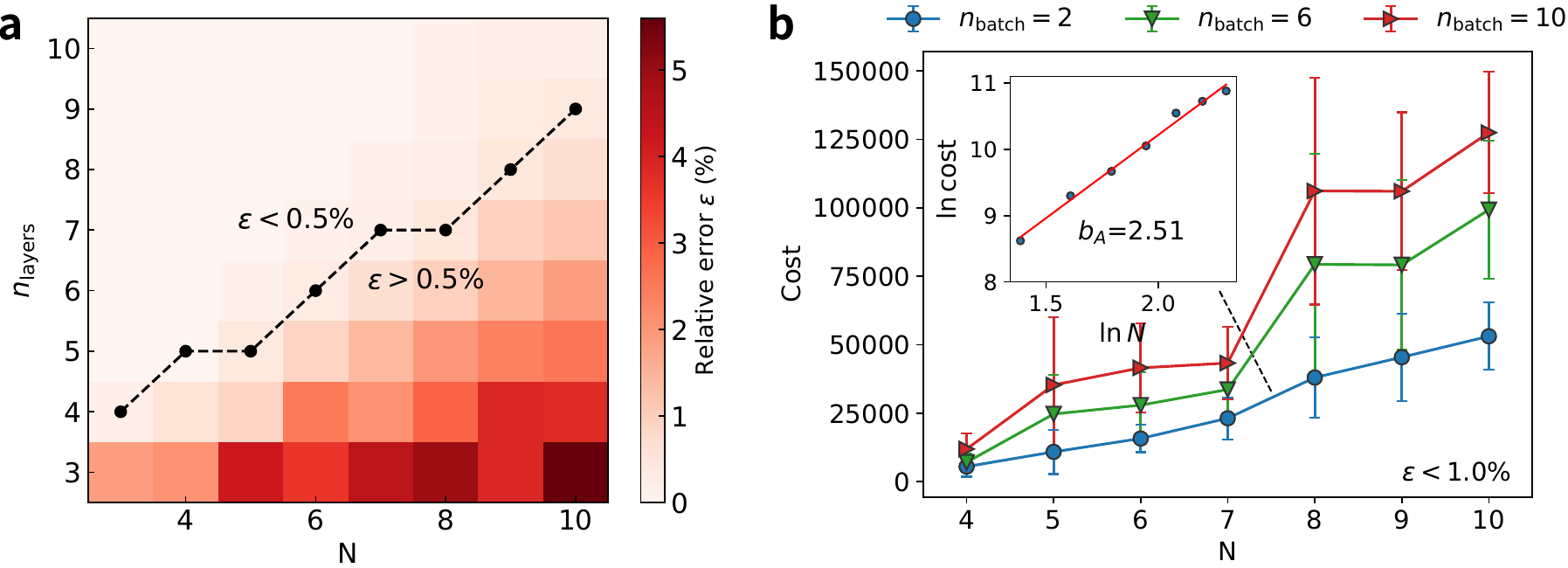}
	\caption{\textbf{Determining the optimal layer number $n_{\text{layer}}$ and the mini-batch size $n_{\text{batch}}$ for the PSR-Adam optimizer}. \textbf{a}, relative error $\epsilon$ with $N$ and $n_{\text{layer}}$ ranging from 3 to 10 after 150 iterations of the full-space PSR-Adam optimization steps. Each point is averaged over 10 trials. The dashed line indicates the minimum value of $n_{\text{layer}}$ that has an averaged error below 0.5\%. The minimum $n_{\text{layer}}$ has a approximately linear scaling in the qubit number. The optimal layer number is determined as $n_{\text{layer}}=5$ for $N=5$ in the experiments, as the circuit representation ability is powerful enough while the circuit is not so deep in order to reduce the decoherence effects. \textbf{b}, computational cost of the sampling PSR-Adam optimization as a function of $N$ with $n_{\text{batch}}=2,\,6,\,10$ with $\epsilon$ less that 1.0\%. The results are averaged over converging runs (iteration steps $n_{\text{iter}}<1000$) from 20 trials. The slope $b_A$ corresponds to the exponent of the growth if the growth was assumed to be polynomial, i.e. $\text{cost}\propto N^{b_A}$. In the inset of \textbf{b}, the $\ln \text{N}$ - $\ln \text{cost}$ plot is close to a straight line.  This indicates that the growth of computational cost follows a polynomial scaling instead of a exponential one, and $n_{\text{batch}}=2$ is the most economic choice for the mini-batch size.}
	\label{fig:ext_fig4}
\end{figure*}

The most important hyper-parameter is the number of layer $n_{\text{layer}}$ which is critical to the expressibility and entangling capability of the quantum circuits.
Using the full-space PSR-Adam optimizer, we study the relationship between the obtained loss function as a function of the layer number $n_{\text{layer}}$ ranging from 3 to 10 in the simulations. The number of iteration $n_{\text{iter}}$ is set to 150, which is beyond the iteration steps required for convergence in the case of the full-space scheme. We plot the relative errors $\epsilon=|\frac{F-F_{exact}}{F_{exact}}|$
for different layer number $n_{\text{layer}}$ and qubit number $N$ in Fig.~\ref{fig:ext_fig4}a. We chose a precision threshold that the relative error is within $0.5\%$. Under the criterion, the scaling of the circuit depth $n_{\text{layer}}$ is approximately linear indicated by the dashed line in  Fig.~\ref{fig:ext_fig4}a. In the experiments, in order to balance the expressibility with the noise and decoherence error, we set $n_{\text{layer}}=5$ in the quantum circuits.

\subsection{Sampling PSR-Adam}
The computational cost for each specific optimizer mainly comes from the evaluation of the gradient (Eq.(\ref{grad1}) and Eq.(\ref{grad2})) of the loss function with respect to the parameter $\theta's$. For the PSR-Adam scheme, the cost is proportional to the number of parameters $n_{\text{para}}$. With the sampling scheme and Eq.(\ref{grad1}), the gradient with respect to $\theta$ takes the following form
\begin{equation}
    	\nabla_{\theta} \mathcal{L}=\nabla_{\theta}\left[\sum^{n_{\text{batch}}}\frac{\left[\beta\bra{\mathbf{x}}U^{\dagger}_{\theta}HU_{\theta}\ket{\mathbf{x}}\right]}{n_{\text{batch}}}\right].
\end{equation}
 A small constant $n_{\text{batch}}$ with respect to the increasing system size gives an exponential reduction to the complexity of the variational algorithm for each iteration since the full-space scheme would require $2^N$ terms in the sum of Eq. (\ref{grad1}). However, the sampling scheme would require a larger number of iterations. 
We consider the total cost for PSR-Adam optimizer with sampling scheme as:
\begin{equation}
\begin{aligned}
     \text{Cost}&=n_{\text{iter}}\times n_{\text{batch}}\times n_{\text{para}}\\
     &=n_{\text{iter}}\times n_{\text{batch}}\times N\times n_{\text{layer}}
\end{aligned}
\end{equation}
with $n_{\text{layer}}$ set according to the contour in  Fig.~\ref{fig:ext_fig4}a. Given a fixed $n_{\text{layer}}$ and $n_{\text{batch}}$, we record the number of iterations $n_{\text{iter}}$ required to reach the precision $\epsilon<1.0\%$. 
Taking into account the randomness that comes from the random initial parameters $\theta$ and the stochastic optimization process, we have used 20 trials for the same $N$ and $n_{\text{batch}}$ and calculate the statistical average of the converging runs, i.e. $n_{\text{iter}}<1000$. The maximal iteration steps 1000 is considered to be large enough since only 1 trial is labeled as 'divergent' among 420 trials in total.
 
We present the results for different $n_{\text{batch}}$ in  Fig~\ref{fig:ext_fig4}b. For the precision threshold and qubit number we considered here, it seems that $n_{\text{batch}}$ can be any integer larger than one. 
This is because the increasing computation complexity in the $n_{\text{batch}}$ sector can be converted to the computation complexity manifested in the larger $n_{\text{iter}}$ value, while maintaining the value of $n_{\text{batch}}$. Thus, it appears to be a trade off between $n_{\text{batch}}$ and the number of iterations $n_{\text{iter}}$. A larger $n_{\text{batch}}$, in general, render a smaller $n_{\text{iter}}$ for the lost function to reach a fixed precision. However, we found that the complexity spared by a smaller $n_{\text{iter}}$ could not balance out the extra cost from larger $n_{\text{batch}}$. As a result, it turned out to be the most efficient choice to set $n_{\text{batch}}=2$ in both numerical and experiment implementation.

In the inset of the  Fig.~\ref{fig:ext_fig4}b, we show the scaling of the total cost with respect to $N$ for $n_{\text{batch}}=2$. We fit the data with power-law $\text{Cost}\propto N^{b_A}$, which is plotted in the $\ln N - \ln \text{Cost}$ coordinate. We observe that the slope in the $\ln N - \ln \text{Cost}$ scatter plot follows a straight line with slope $b_A=2.51$, indicating a polynomial scaling.

\subsection{Sampling SPSA-Adam}
Here we consider the case of the SPSA-Adam optimizer, where the evaluations of the gradient with respect to all parameters are replaced by simultaneous shifts in all parameters averaged over $n_{\text{SPSA}}$ times. Thus the cost is proportional to $n_{\text{SPSA}}$ and can be estimated as 
\begin{equation}
\text{Cost}=n_{\text{iter}}\times n_{\text{batch}}\times n_{\text{SPSA}}
\end{equation}

We set the mini-batch size $n_{\text{batch}}$ to 2, as determined in the last subsection, and investigate the scaling of computational cost via $N$. The SPSA average number $n_{\text{SPSA}}$ is selected from \{1,3,5,7,9,11,13\} and each set of hyperparameters was trialed 20 times. We record the averaged number of iterations $n_{\text{iter}}$ required for each value of N to reach the precision $\epsilon<3.0\%$. The maximum $n_{\text{iter}}$ is set to 1500. Any trial with $n_{\text{iter}}$ exceeded the maximal $n_{\text{iter}}$ is considered divergent, and the corresponding parameter $n_{\text{SPSA}}$ is considered to be too small for the corresponding N.

\begin{figure}[t]
	\centering
	\includegraphics[width=8.2cm]{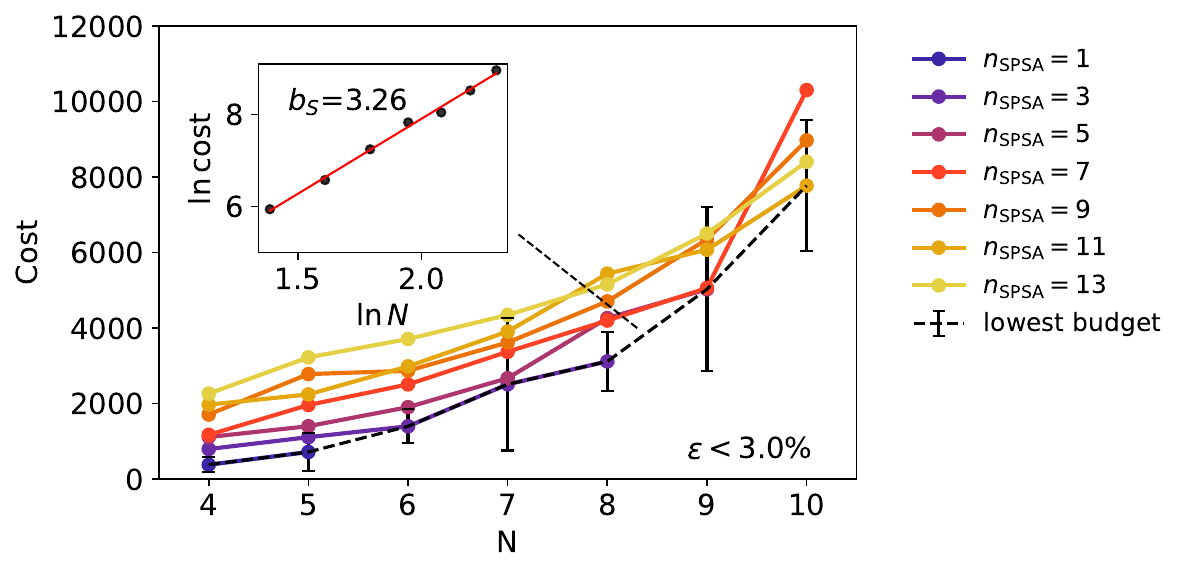}
	\caption{\textbf{Scaling of the computational cost for the SPSA-Adam optimizer with an increasing qubit number $\mathbf{N}$}. Computational cost with increasing $N$ for $n_{\text{SPSA}}=1,\,3,\,5,\,7,\,9,\,11,\,13$ with $\epsilon$ less than 3.0\%. Given a fixed precision, there is a threshold for $n_{\text{SPSA}}$ for each $N$ and therefore the curves corresponding to different $n_{\text{SPSA}}$ terminate at certain values of $N$. The dashed lines indicate the minimal cost simulated for each $N$. The results are averaged over converging runs (iteration steps $n_{\text{iter}}<1500$) from 20 trials. In the inset, the $\ln \text{N}$-$\ln \text{cost}$ plot of the minimum cost is close to a straight line and the slope, i.e. $b_S=3.26$ corresponds to the exponent of the growth if the growth was assumed to be polynomial, i.e. $\text{cost}\propto N^{b_S}$. This indicates that the growth of computational cost follows a polynomial scaling instead of a exponential one.}
	
	\label{fig:ext_fig5}
\end{figure}

Similar to the relation between $n_{\text{batch}}$ and $n_{\text{iter}}$, there appears to be a trade off where $n_{\text{iter}}$ tends to decrease as $n_{\text{SPSA}}$ increases. 
The parameter $n_{\text{SPSA}}$, however, does have a different scaling behavior from $n_{\text{batch}}$, where we found that $n_{\text{batch}}=2$ is always the most efficient option. Starting from $N=6$ with $n_{\text{SPSA}}=1$, $N=9$ with $n_{\text{SPSA}}=3$, and $N=10$ with $n_{\text{SPSA}}=5$, there are cases where the value of $n_{\text{iter}}$ surpassed the maximum iteration limit $n_{\text{iter}}=1500$. As a consequence, those data were not calculated and missing in  Fig.~\ref{fig:ext_fig5}, as we consider those value of $n_{\text{SPSA}}$ are below the minimum threshold for the corresponding qubit number.
The scaling of computational cost is shown in  Fig.~\ref{fig:ext_fig5}, where the dashed line represents the minimum cost required for each $N$ with $\epsilon<3.0\%$. The scaling of the minimum cost is also plotted in a $\ln N$-$\ln \text{cost}$ in the inset. It again illustrates a power-law behavior $\text{cost}\propto N^{b_S}$ with $b_S=3.26$.
As shown in Fig.~\ref{fig:ext_fig4}b and Fig.~\ref{fig:ext_fig5}, the SPSA-Adam optimizer has a scaling of higher-order compared with the PSR-ADAM optimizer even with lower precision. The result agrees with theoretical analysis that the zero's order or gradient approximating methods have higher-order scaling in terms of both increasing qubit number and precision, compared with optimizations using analytic gradient measurements~\cite{harrow2021low-depth}. We therefore only investigated the scaling result of SPSA-Adam optimization for $\epsilon<3.0\%$ but not for higher precision due to limited computational resources.

However, the SPSA-Adam optimizer is shown to be more efficient than the PSR-Adam optimizer in this experiment as the former only takes a few (in this experiments $n_{\text{SPSA}}=10$) averages compared with $n_{\text{para}}=25$ parameters for PSR-Adams optimizer. Moreover, the SPSA-Adam optimizer will always hold the advantage when certain symmetry, e.g. parity or translation symmetry, is implemented in the circuit, as the same parameter applies to different quantum gates. The parameter shift method would not benefit much from symmetry implementation whereas the SPSA method for calculating the gradient would reduce a considerable amount of cost in such cases. On top of that, parameter shift rules have their limitations especially when dealing with gates that do not square to a constant time the identity, while the SPSA method will not be limited by the choice of quantum circuits ansatz, thus has an advantage over the PSR-Adam optimizer.

\section{Variance and the self-verifiable feature}\label{app_F}
The loss function used in the optimization is the variational free energy
\begin{equation}
	\mathcal{L}=\mathbb{E}_{x \sim p_{\phi} (\x)} [\ln p_{\phi} (\x)/{\beta} + 
	\langle \psi_{\mathbf{\theta}}(\x)|\hat{H} |\psi_{\mathbf{\theta}}(\x)\rangle],
\end{equation}
which is evaluated as a mean value over the variational distribution.
Its variance usually offers additional information about the confidence of optimizations. This is based on the fact that the variance is zero when the learning is exact. To illustrate this, suppose the optimization is perfect such that $\mathcal L$ equals the exact free energy $F$ and $p_\phi(\x)$ characterizes exactly the level statistic $P(\x)$, with 
$$P(\x) = \frac{e^{-E_\theta(\x)}}{e^{-\beta F}},
$$
with energy computed as 
$$E_\theta(\x)=\langle \psi_{\mathbf{\theta}}(\x)|\hat{H} |\psi_{\mathbf{\theta}}(\x)\rangle.$$
Then we can see that 
$$\ln p_{\phi} (\x)/{\beta} +\langle \psi_{\mathbf{\theta}}(\x)|\hat{H} |\psi_{\mathbf{\theta}}(\x)\rangle=F,$$ which is independent of $\x$, indicating that the variance of $\langle \psi_{\mathbf{\theta}}(\x)|\hat{H} |\psi_{\mathbf{\theta}}(\x)\rangle$ is zero.
However, when we encountered a zero-variance during learning which produces the zero gradients of the loss funciton with respect to the model parameters, it does not indicate that the learning is exact. To see this, suppose $$\ln p_{\phi} (\x)/{\beta} +\langle \psi_{\mathbf{\theta}}(\x)|\hat{H} |\psi_{\mathbf{\theta}}(\x)\rangle = \textbf {C},$$ with $C$ denoting a constant, one has 
\begin{equation}
p_{\phi}(\x) = e^{\beta C}e^{ -\beta \langle\psi_{\mathbf{\theta}} (\x)|\hat{H} |\psi_{\mathbf{\theta}}(\x)\rangle}
.
\end{equation}
This means that when the zero-variance condition is reached, the learned distribution is a Boltzmann distribution. However notice that $e^{\beta C}$ does not necessarily be $e^{\beta F}$, resulting in a Boltzmann distribution that deviates from the true distribution. This is usually due to the \textit{model collapse} effect that the variational ansatz can only explore part of the configuration space. 

Nevertheless, in practice, the amount of variance can be used to justify and verify the quality of learning. Empirically, a small variance usually indicates a small variability of model prediction, yielding a small gap between the variational free energy and the exact free energy, e.g. in the $\beta$-VQE~\cite{liu_solving_2021} algorithm and its classical counter part~\cite{wu2019solving}.

\section{Computing entropy of the classical distribution}\label{app_G}
The advantage of our variational framework is that the entropy can be computed efficiently with a polynomial computational complexity in the number of variables, by designing the classical distribution.
In this work, we have employed the ansatz of the product distribution 
\begin{equation}
	p_{\phi} (\x) = \prod_i p_{\phi_i} (x_i) = \prod_i \phi_i^{x_i} (1 - \phi_i)^{1 - x_i}.
\end{equation}
its entropy can be estimated analytically
\begin{align}
S=-\sum_i\sum_{x_i}p_{\phi_i}(x_i)\log(p_{\phi_i}(x_i)).
\end{align}
which does not introduce statistical error. The
ansatz can be extended to a more expressive distribution with entropy computed using samples. For example, one could parameterize the joint distribution as a product of conditional probabilities
\begin{equation}
p_{\phi}(\x) = \prod_ip(x_i|x_1,\cdots,x_{i-1}).
\end{equation}
Here we need to pre-determine an order for the variables $\x$.
This is actually the chain rule of probabilities, also known as the ~\textit{autoregressive} model~\cite{DLbook}. As a simple example, let us consider a simple example with $4$ binary variables 
\begin{align}
    &p_{\phi}(x_1,x_2,x_3,x_4) \nonumber\\
    &=p(x_4|x_1,x_2,x_3)p(x_1,x_2,x_3)\nonumber\\
    &=p(x_4|x_1,x_2,x_3)p(x_3|x_1,x_2)p(x_1,x_2)\nonumber\\
    &=p(x_4|x_1,x_2,x_3)p(x_3|x_1,x_2)p(x_2|x_1)p(x_1).
\end{align}
At the first glance, some of the conditional probabilities are still difficult to express due to an exponential number of parameters to the number of variables in it. In practice, one can adopt an efficient model such as neural networks with a polynomial number of parameters to the number of variables for parameterizing the conditional probability.
There are two essential properties of this representation, the ability to compute normalized probability $p_\phi(\x)$ for any configuration $\x$, and a fast sampling algorithm.
The first property is obvious because each conditional probability is normalized, as
\begin{equation}    
\sum_{x_i}p(x_i|x_1,\cdots,x_{i-1}) = 1,
\end{equation}
which means that $p_\phi(\x)$ is naturally normalized.
The second property is known as \textit{ancestral sampling}~\cite{bishop2006pattern}, which samples variables one by one given that we have stored every normalized conditional probability. Again taking the example with $4$ variables, we can first toss a coin to fix a configuration for $x_1$ according to $p(x_1)$, then toss a coin to fix a configuration for $x_2$ according to $p(x_2|x_1)$, and configurations for $x_3$ and $x_4$ in turn. Moreover, a large number of unbiased samples can be drawn in this way parallelly.

Equipped with the properties, one can easily make an accurate estimate of entropy using
\begin{align}
    S &= -\sum_\x P_\phi(\x)\log(P_\phi(x))\\
    & = -\mathbb E_{x\sim P_\phi}\log(P_\phi(x))\nonumber\\
    & \approx -\frac{1}{m}\sum_{\mu=1}^m \log(P_\phi(x_\mu)),
\end{align}
where in the last equation we have used $m$ samples to approximately estimate the expectation value.

\begin{figure*}[t]
	\includegraphics[width=15cm]{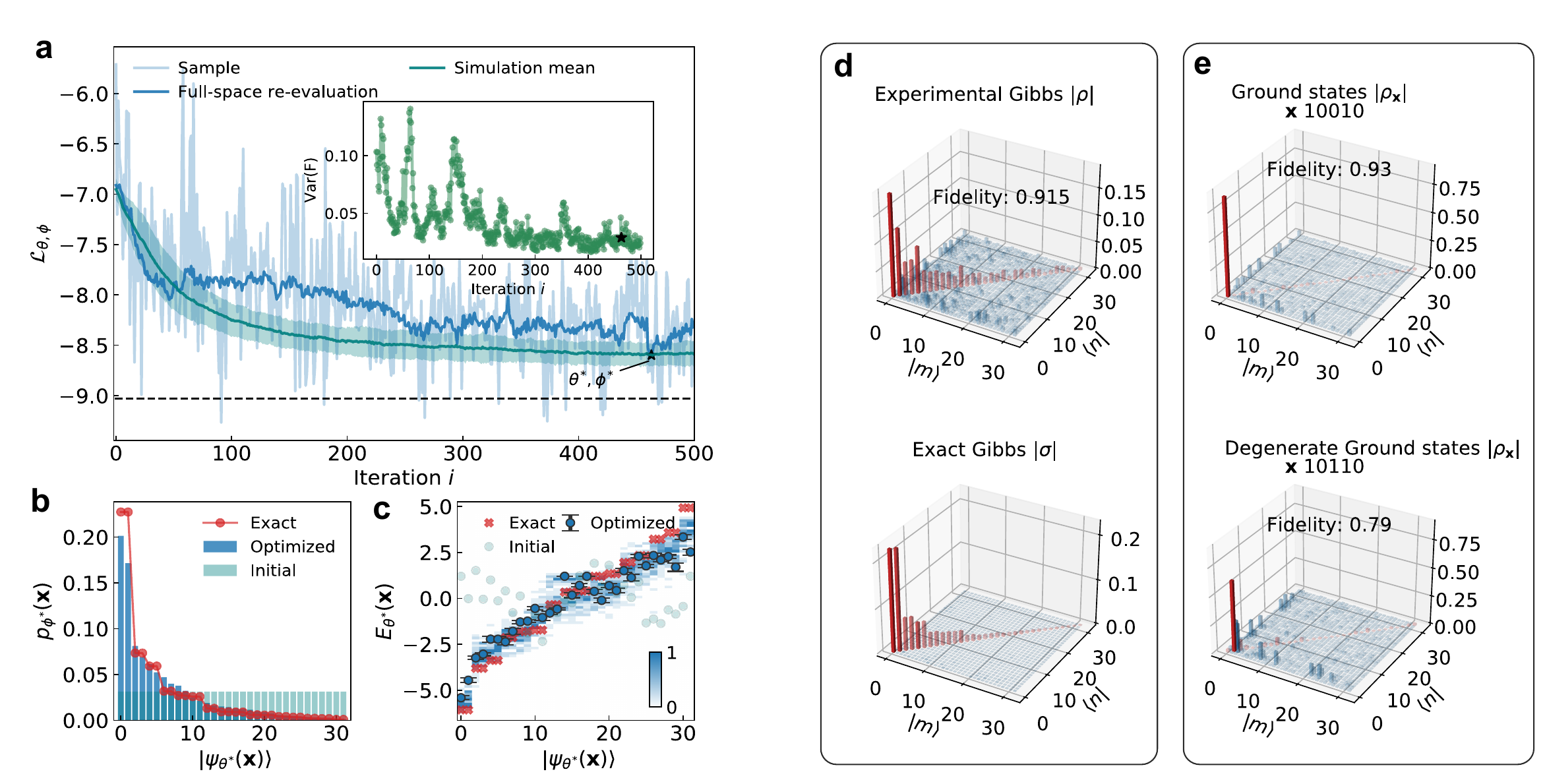}
 	\caption{ \textbf{Experimental optimization results for quantum XXZ model with $h=0$, $\Delta=0.3$, and $\beta=0.5$}. \textbf{a}, optimization trajectory (light blue) and full-space re-evaluation results (blue) of the loss function with the sampling scheme and the SPSA-Adam optimizer. The shaded area denotes the standard error of 50 full-space re-evaluations with the mean (teal) at the center. The black dashed line is the exact value of free energy. The inset shows a self-verify indicator by tracking the sample variance of the loss function. \textbf{b} and \textbf{c}, optimized probability distribution $p_{\phi^*}(\x)$ (\textbf{b}) of the Gibbs ensemble and eigenenergy $E_{\theta^*}(\x)$ (\textbf{c}) of the target Hamiltonian with parameters $\theta^{*}, \phi^{*}$ corresponding to the iterative step marked by the star in \textbf{a}. The energy expectation are sorted by probabilities. The density plot of eigen energies in \textbf{c} are obtained from 50 numerical simulations.  \textbf{d} and \textbf{e}, density matrix of the Gibbs states and eigenstates obtained experimentally for the quantum XXZ model. \textbf{d}, the density matrix of the optimized Gibbs states $\hat{\rho}^*$ (upper panel) compared with the exact Gibbs states (lower panel). \textbf{e}, density matrix $\rho_{\x}$ of prepared specified eigenstates $|\psi_{\theta^*}(\x)\rangle$, where the corresponding exact eigenstate $|n\rangle$ is identified by $n$th probabilities $p_{\phi^*}(\x)$ sorted in a descending order. In all plots, only the amplitudes of the density matrix are shown. The diagonal elements and the off diagonal elements are colored red and blue respectively.
             }
	\label{fig:ext_fig1}
\end{figure*} 

\begin{figure*}[t]
	\includegraphics[width=15cm]{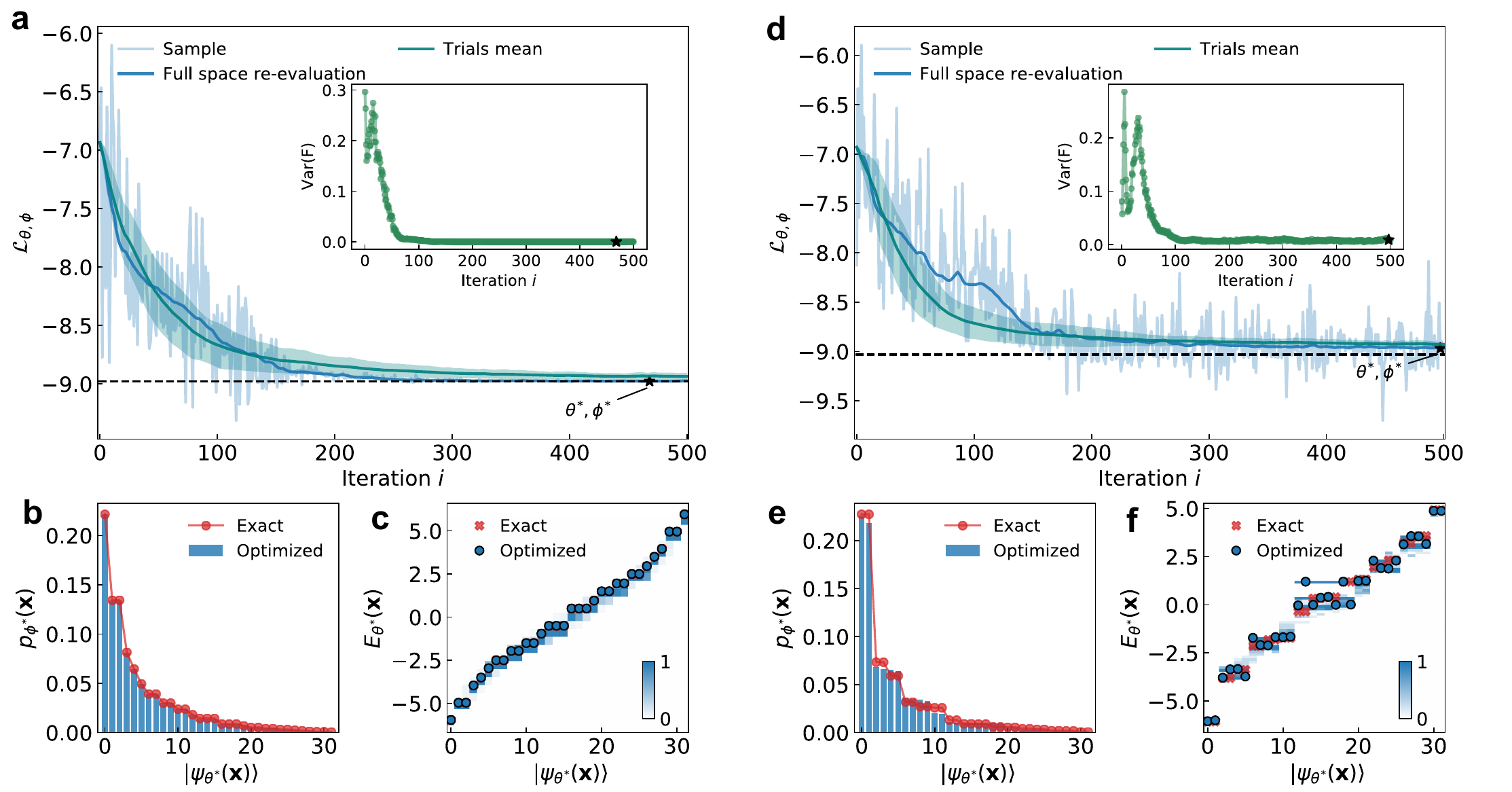}
	\caption{ \textbf{Ideal numerical optimization results.}
	\textbf{a}, \textbf{b}, and \textbf{c}, results for the 
	quantum XY model with $h=0.5$, $\beta=0.5$. \textbf{a}, optimization trajectory (shallow blue) and full-space re-evaluation results (blue) of the loss function  with the sampling scheme and the SPSA-Adam optimizer. The shaded area denotes the standard error of 50 ideal numerical trials of the full-space re-evaluated optimizations, with the mean (teal) at the center. The black dashed line is the exact value of  the free energy. The inset shows a self-verify indicator by tracking the sample variance of the loss function. \textbf{b} and \textbf{c}, optimized probability distribution $p_{\phi^*}(\x)$ (\textbf{b}) of the Gibbs ensemble and the eigen energy $E_{\theta^*}(\x)$ (\textbf{c}) of the target Hamiltonian with parameters $\theta^{*}, \phi^{*}$ corresponding to the iterative step marked by the star in \textbf{a}. The energy expectations are sorted by probabilities and show a one-to-one correspondence with the exact value. The density plot of the eigen energies in \textbf{c} are obtained from 50 ideal trails.  \textbf{d}, \textbf{e},  and \textbf{f}, results for the quantum XXZ model with $h=0$, $\Delta=0.3$, and $\beta=0.5$. Due to the limited expressive power of our variational ansatz for the XXZ model, some eigen energies identified by probabilities deviate from the exact values, which are also shown in  Fig.~\ref{fig:ext_fig3}}
	\label{fig:ext_fig2}
\end{figure*}

\section{Additional experimental and numerical results}\label{app_H}
\begin{figure}[t]
	\includegraphics[width=9cm]{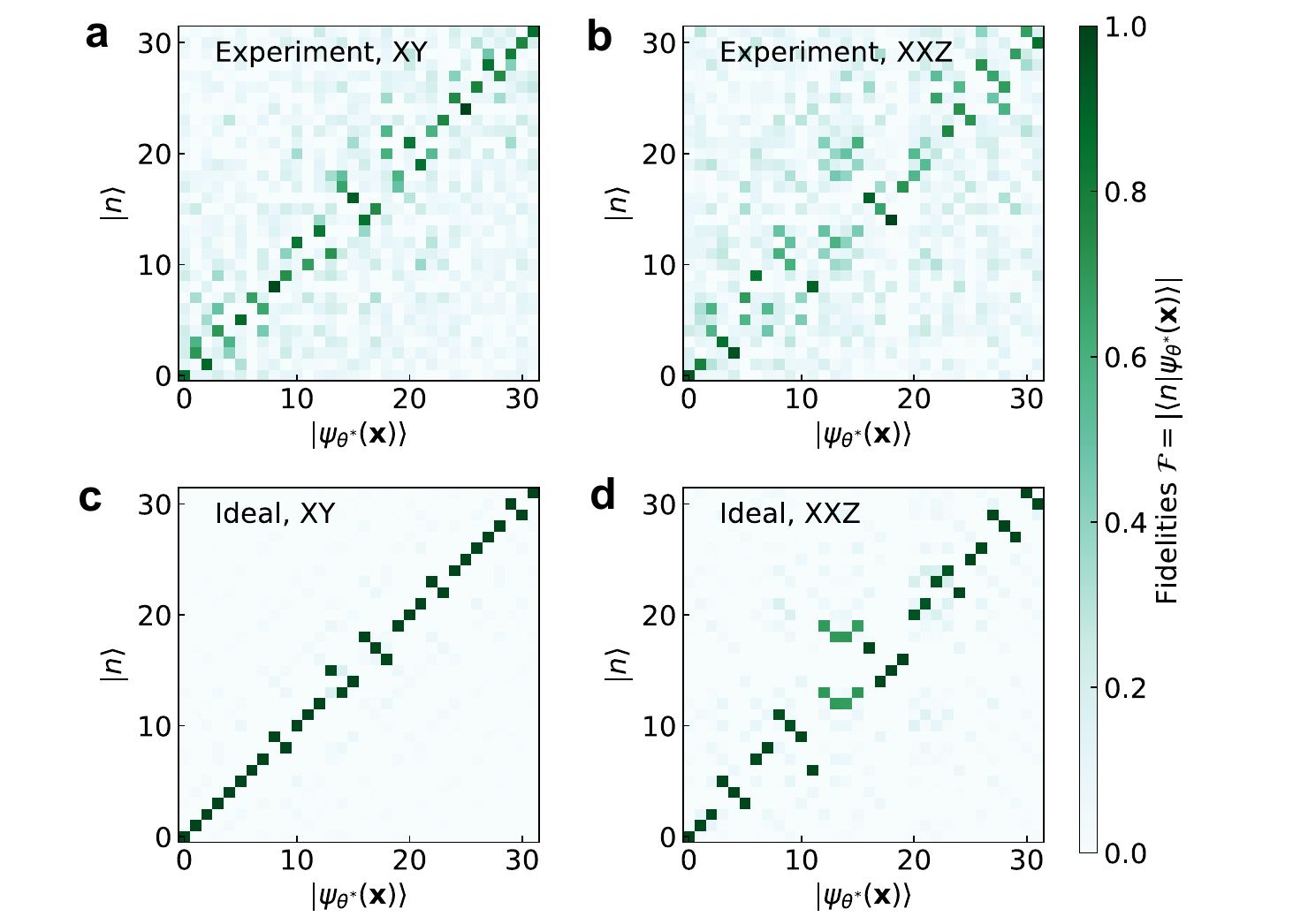}
	\caption{ \textbf{The fidelity matrix between the optimized eigenstates $|\psi_{\theta^{*}}(\mathbf{x})\rangle$ and the exact eigenstates $|n\rangle$.} \textbf{a} and \textbf{b}, the fidelity matrix for the quantum XY and XXZ model obtained in experiments. The matrix elements are concentrated near the diagonal entries, indicating that the eigenstates of the target Hamiltonian are prepared accurately in the experiments. \textbf{c} and \textbf{d}, The fidelity matrix from the ideal simulations for the quantum XY and XXZ model respectively. In the XY case, most matrix elements are exactly diagonal-located while the off-diagonal elements represent the degeneracy of the target energy spectrum. For the XXZ model, the elements that deviated from the diagonal entries are given by both the degeneracies and the in-exact correspondence between the probability-identified eigen energy and the exact eigen energy. 
             }
	\label{fig:ext_fig3}
\end{figure}

In this Appendix we present the experimental optimization results for XXZ models (Fig.~\ref{fig:ext_fig1}), ideal numerical simulation results for XY and XXZ models (Fig.~\ref{fig:ext_fig2}) and experimental obtained fidelity matrix for XY and XXZ models (Fig.~\ref{fig:ext_fig3}).

\clearpage

\end{document}